\documentclass[12pt]{article}
\usepackage{epsfig}
\usepackage{epstopdf}
\usepackage{float}
\usepackage{indentfirst}
\usepackage{multirow}
\usepackage{amsmath}
\usepackage{fullpage}
\usepackage{booktabs}
\usepackage[utf8]{inputenc}
\usepackage{float}
\usepackage{listings}
\usepackage{xcolor}
\usepackage{changes}
\usepackage{caption}
\usepackage{subcaption}
\usepackage{amsmath}
\usepackage{lscape}
\usepackage{amsfonts}
\usepackage[backend=biber, style=authoryear]{biblatex}
\title{Conditional-Marginal Nonparametric Estimation for Stage Waiting Times from Multi-Stage Models under Dependent Right Censoring}
\author{
Xiaoxi Zhang, Peihua Qiu, and Somnath Datta\thanks{email: somnath.datta@ufl.edu} \\
Department of Biostatistics, University of Florida, Gainesville, FL 32611, USA
}
\date{\today}

\begin{document}

\maketitle

\begin{abstract}

We investigate two population-level quantities (corresponding to complete data) related to uncensored stage waiting times in a progressive multi-stage model, conditional on a prior stage visit. We show how to estimate these quantities consistently using right-censored data.
The first quantity is the stage waiting time distribution (survival function), representing the proportion of individuals who remain in stage 
$j$ within time $t$ after entering stage $j$. The second quantity is the cumulative incidence function, representing the proportion of individuals who transition from stage $j$ to stage $j'$ within time $t$ after entering stage $j$. To estimate these quantities, we present two nonparametric approaches. The first uses an inverse probability of censoring weighting (IPCW) method, which reweights the counting processes and the number of individuals at risk (the at-risk set) to address dependent right censoring. The second method utilizes the notion of fractional observations (FRE) that modifies the at-risk set by incorporating probabilities of individuals (who might have been censored in a prior stage) eventually entering the stage of interest in the uncensored or full data experiment. Neither approach is limited to the assumption of independent censoring or Markovian multi-stage frameworks. Simulation studies demonstrate satisfactory performance for both sets of estimators, though the IPCW estimator generally outperforms the FRE estimator in the setups considered in our simulations. These estimations are further illustrated through applications to two real-world datasets: one from patients undergoing bone marrow transplants and the other from patients diagnosed with breast cancer.

\end{abstract}

\section{Introduction}\label{intro}
Traditional survival analysis follows a straightforward framework in which individuals irreversibly transition from one stage (e.g., alive) to another (e.g., death). However, certain diseases, which involve multiple events or stages, are too complex to be described by such binary characteristics. A multi-stage model offers a more effective approach to capturing these complex multivariate survival scenarios. In a multi-stage model, individuals progress through a succession of stages, each representing their current disease condition. Within the context of a multi-stage model, similar to the survival function and hazard rate function in traditional survival analysis, the quantities of interest at each of the stages are marginal functions such as the stage occupation probabilities, cumulative incidence function, entry/exit time distributions, and even the transitions probabilities between stages (generally, under the assumption of a Markov model). 

In this paper, we aim to use right-censored data from a progressive multi-stage model to estimate the following two quantities related to stage waiting times conditional on a prior stage visit: (1) stage waiting time distribution: the proportion of individuals who enter stage $j$ and stay in stage $j$ within waiting time $t$ of entering stage $j$, and (2) cumulative incidence function: the proportion of individuals who enter stage $j$ and subsequently leave stage $j$ for stage $j'$ within waiting time $t$ of entering stage $j$.

\begin{figure}[htb]
\centering
\includegraphics[scale=.50]{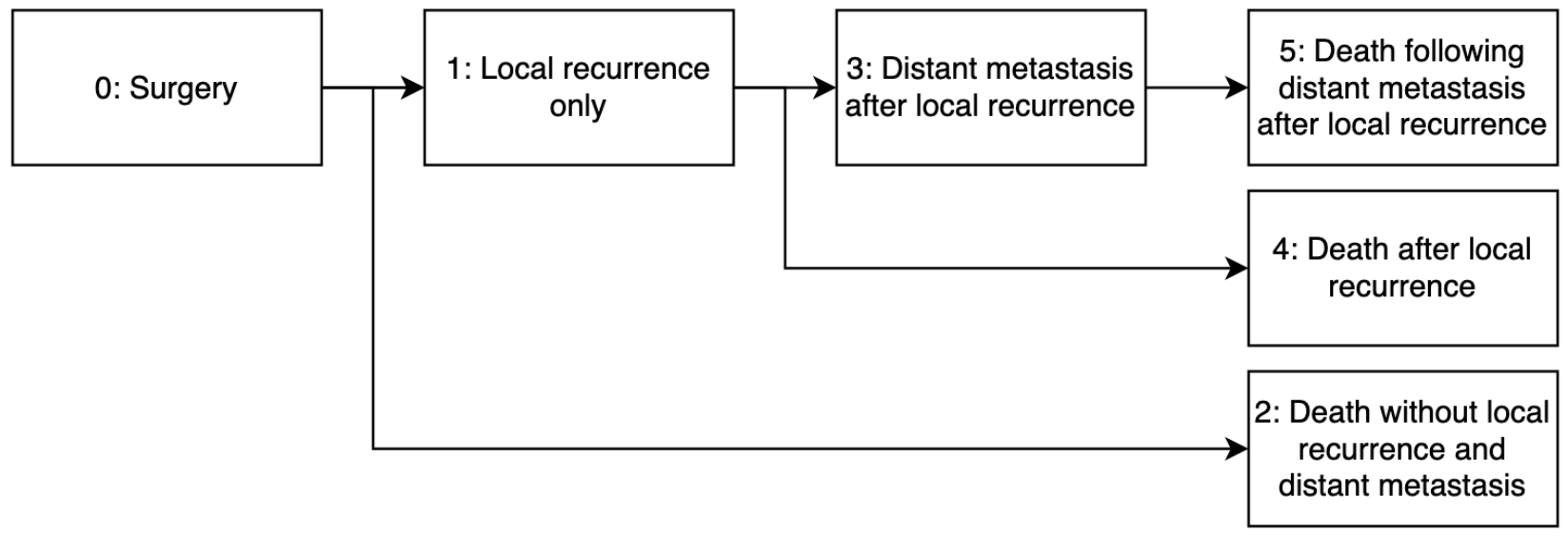} 
\caption{A simplified six-stage progressive survival model for breast cancer data.}\label{fig:1-1}
\end{figure}

A six-stage survival model for breast cancer, shown in Figure \ref{fig:1-1}, is employed to illustrate the two quantities of interest. It should be noted that this multi-stage model depicted in Figure \ref{fig:1-1} serves as a simplified example to introduce key concepts and explain our methods and simulation studies presented later. A more detailed multi-stage model for breast cancer is provided in Web Appendix D of Supplementary Material. As described in Figure \ref{fig:1-1}, individuals may move through different health stages over time. The stages are defined by three post-surgery events: the onset of local recurrence, the onset of distant metastasis, and death. Beginning in the surgery stage (stage 0), which represents individuals who have just undergone surgery and remain event-free, individuals may either die without experiencing any post-surgery events (stage 2) or develop local recurrence (stage 1). After entering stage 1, individuals may either die (stage 4) or progress to both distant metastasis and local recurrence (stage 3). Individuals in stage 3 may subsequently die (stage 5). The focus of our study is to estimate: (1) the probability that individuals remain in stage 3 within time $t$ after entering stage 3, given a prior visit to stage 1, and (2) the probability that individuals transition from stage 3 to stage 5 within time $t$ after entering stage 3, given a prior visit to stage 1. Stages 2, 4, and 5 are defined as terminal stages since no events can happen after death.

The aforementioned conditional probabilities in multi-stage models have historically been understudied, primarily because multi-stage systems are often assumed to follow the Markov assumption, where the future stage and transition time depend only on the current stage and not on the past stages (and event times). However, there is increasing recognition that the Markov assumption can be problematic since it may be difficult to verify and often be violated in real-world applications. For example, in complex disease progressive multi-stage models, an individual’s future health stage may be influenced by the entire sequence of prior stages (and those entry times), not just the most recent one (and the calendar time). As far as we know, few existing methods and analyses in the literature make additional structural assumptions, such as Markovity or semi-Markovity (i.e., Datta and Satten, 2002), and/or parametric model assumptions.  It is important to emphasize that our estimators do not rely on these assumptions, which allows them to be applicable to a broader range of progressive multi-stage data.

In most multi-stage analyses, formulation of quantities such as stage occupation probabilities, cumulative incidence function, entry/exit time distribution, and transition probability matrix is typically based on calendar time (e.g., Satten et al., 2001; Datta and Satten, 2002; Yang et al., 2023). Under the calendar time framework, time is measured from the start of follow-up to the occurrence of all events across all different stages. However, our estimation is based on waiting time, which is measured from the start of follow-up to the occurrence of events in the specific stage of interest. The concept of waiting time has been discussed in the literature (e.g., Lin et al., 1999; Wang and Wells, 1998; Lan and Datta, 2010). Compared to calendar time, waiting time may have greater clinical significance. In the example of the six-stage progressive survival model (i.e., Figure \ref{fig:1-1}), when evaluating the effect of a treatment on this episodic disease, it is often necessary to assess whether the treatment delays the time from stage 1 to stage 3, as well as the time from stage 3 to stage 5, and so on. The calendar time for the initiation of the treatment to stage 5 is less clinically relevant since delaying stage 3 will inevitably lengthen the calendar time to stage 5, even if the treatment becomes ineffective after stage 3.

Another important issue we aim to address while dealing with multi-stage models is dependent right censoring. Using the six-stage survival model (i.e., Figure \ref{fig:1-1}) as an example, the longer an individual takes to transition from stage 1 to stage 3, the higher the probability that the individual would be censored after entering stage 3. Robins and Rotnitzky (1992) proposed using inverse probability of censoring weights (IPCW) to address this dependent censoring mechanism in the traditional two-stage survival analysis problem. Satten et al. (2001) extended the IPCW approach by applying Aalen’s linear model for the cumulative hazard of being censored. Weighted estimation in various multi-stage contexts has been explored in several recent studies (e.g., Datta and Satten, 2002; Lan and Datta, 2010; Mostajabi and Datta, 2013). However, applying the IPCW approach to the nonparametric estimation of stage waiting-time quantities conditional on past stages in multi-stage data is not a straightforward task and has not been thoroughly explored in previous work. In this paper, we demonstrate the efficiency of the IPCW approach in addressing this challenge theoretically and numerically. Yang et al. (2023) incorporated the concept of fractional observation to construct the at-risk set, accounting for the probability of patients censored in past stages eventually entering future stages of interest.
However, incorporating fractional observation into the estimation of waiting-time-related quantities in multi-stage models is a complex task that has not been explored before. In this paper, we address this gap by using the idea of a ``unique path'' when constructing the at-risk set.

Thus, to summarize, our methods have the following nuances: (i) unlike most papers, they are formulated in terms of state waiting times instead of event calendar times (e.g., state entry/exit times), (ii) we allow for non-Markov or non-semi Markov systems, (iii) the censoring mechanism under play may depend on internal (or external) covariates such as the currently occupied stage, (iv) instead of considering the entire population, inquiries are restricted to individuals who pass through a certain stage of interest in the full data experiment.

The remainder of the paper is organized as follows: Section \ref{methods} provides a detailed description of our proposed estimators, supplemented by an illustrative example. Section \ref{simulation} presents simulation results evaluating the performance of these estimators. Section \ref{case studies} demonstrates the application of the proposed methods through two real-data examples. The paper concludes with several remarks in Section \ref{concluding remarks}. Supplementary Material includes an outline of the proof of the main theorem, additional numerical results and a real case study. Supplementary material is included at the end of this paper.

\section{Methods}\label{methods}

In this section, we first introduce the notations and then utilize the illustrative example of a six-stage survival model (i.e., Figure \ref{fig:1-1}) to elucidate the nonparametric estimation techniques through two different methods: the inverse probability of censoring weighted (IPCW) estimator and the fractional risk estimator (FRE). Subsequently, we present the formulas for the nonparametric estimators in a general multi-stage model.

Consistent with some other papers (e.g., Lan and Datta, 2010; Yang et al, 2023; Anyaso-Samuel et al, 2023), our multi-stage model does not rely on the Markov (e.g., the hazard rates of transition are functions of calendar times) or semi-Markov assumptions (the hazard rates of transition are functions of waiting times). This paper only focuses on a progressive tree structure, defined as an acyclic, hierarchical, and directed graph where nodes represent sequential stages and edges denote stepwise transitions between two stages, for two main reasons. First, defining stage waiting times in multi-stage models is challenging when stages can be entered multiple times. In this case, it becomes necessary to distinguish between stage waiting times for the first entry, second entry, and so on. Second, estimating quantities related to stage waiting times, conditional on prior stage visits, is feasible in such a progressive tree structure, where individuals in the same stage have the same histories of prior stage visits (although not the state entry/exit times).

\begin{figure}[htb]
\centering
\includegraphics[scale=.60]{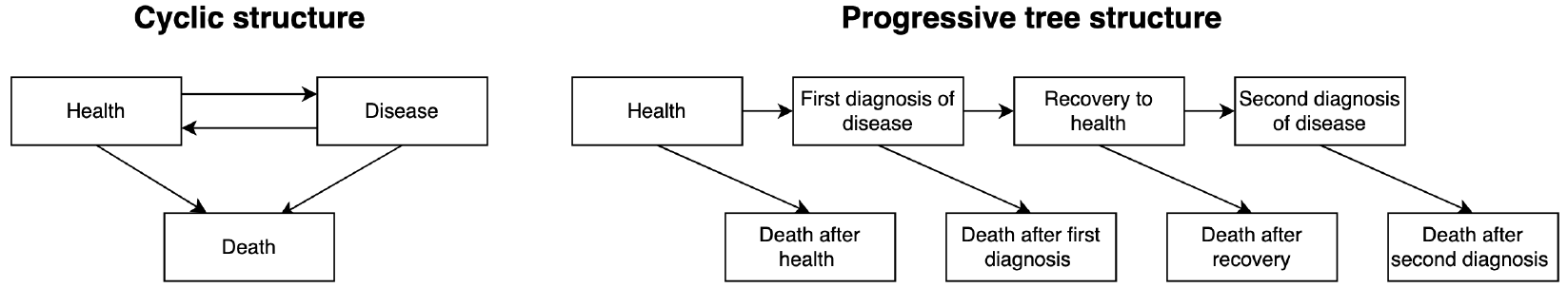} 
\caption{Transformation of an illness-death model from a cyclic structure to a progressive tree structure.}\label{fig: illness_death_model}
\end{figure}

We note that even a complex multi-stage model involving repeated events, such as cyclic structures where transitions happen back and forth between stages, can be expanded into a progressive tree structure by dividing a single stage in the original model into multiple distinct stages in the progressive tree structure. An example is illustrated in Figure \ref{fig: illness_death_model}, which shows the transformation of the illness-death model from a cyclic structure to a progressive tree structure. The left panel of Figure \ref{fig: illness_death_model} depicts a cyclic structure, where individuals begin in the ``health'' stage and transition back and forth between ``health'' and ``disease" and may enter the terminal stage ``death'' from either. This cyclic structure is converted into the progressive tree structure shown in the right panel of Figure \ref{fig: illness_death_model} through the following two steps: (1) the cyclic transitions between ``health" and ``disease" are replaced with a chain of distinct stages — ``health", ``first diagnosis of disease", ``recovery to health", and ``second diagnosis of disease" etc., and (2) the ``death" stage is divided into four (plus) separate stages: ``death after disease", ``death after first diagnosis", ``death after recovery", and ``death after second diagnosis". However, it should be noted that our proposed estimations are based on this expanded progressive structure rather than the original structure. This may or may not be possible for all applications. We touch upon this issue further in the discussion section.

\subsection{Notation and definition}\label{method: def}
Table \ref{tab: notation} shows the notations used in the multi-stage model. The data are subject to right censoring, and hence the values of $T_{ij}^{*}$, $U_{ij}^{*}$, $T_{i}^{*}$, and $X^*_{ij}$ may not always be observed. We assume the data consists of i.i.d. replicates of $\left\{T_{ij}, U_{ij}, \delta_{ij}, \gamma_{ij}, 1\le i \le n, 1\le j \le J \right\}$.

\begin{table}[htb]
  \scriptsize
  \centering		
  \setlength{\belowcaptionskip}{+0.2cm}
	\caption{Notations for a general $J$-stage survival model.}
  \label{tab: 4-1}
	\begin{tabular}{cc} 
		\hline
		Symbol &Meaning \\
		\hline	
		$J$ & Number of stages, $1\le j \le J$ \\  
        $n$ & Number of individuals, $1\le i \le n$ \\  
        $T_{ij}^{*}$ & Time the $i$th individual enters stage $j$ ($=\infty$ if the $i$th individual never enters stage $j$) \\  
        $U_{ij}^{*}$ & Time the $i$th individual leaves stage $j$ ($=\infty$ if the $i$th individual never leaves stage $j$) \\  
        $T_{i}^{*}$ & $\max_j \{ T_{ij}^{*}|T_{ij}^{*}<\infty \}$: Time of $i$th individual's last transition \\  
        $X^*_{ij}$ & $I(T_{ij}^{*}<\infty)$: Indicator function that takes value 1 if the $i$th individual ever enters stage $j$ and 0 otherwise. \\  
        \hline
        $C_{i}$ & Censoring time for the $i$th individual \\  
        $\gamma_{ij}$ & $I(T_{ij}^{*}\le C_i)$: Indicator function that takes value 1 if the $i$th individual is observed to enter stage $j$ and 0 otherwise.  \\  
        $\delta_{ij}$ & $I(U_{ij}^{*}\le C_i)$: Indicator function that takes value 1 if the $i$th individual is observed to leave stage $j$ and 0 otherwise. \\  
        $T_i$ & $\min (T_i^*,C_i)$\\
        $T_{ij}$ & $\min (T_{ij}^*,C_i)$\\
        $U_{ij}$ & $\min (U_{ij}^*,C_i)$\\
        $\delta_i$ & $I (T_i^*\le C_i)$\\
		\hline
	\end{tabular}	
   \label{tab: notation}
\end{table}

Before extending our estimators to general multi-stage models, we begin with the illustrative toy example of the six-stage survival model (i.e., Figure \ref{fig:1-1}) to define the quantities of interest and provide estimators of these quantities for uncensored data for a better illustration. This six-stage survival model has been thoroughly discussed in Section \ref{intro}. We aim to estimate the following two quantities: (1) stage waiting time distribution of stage 3 conditional on the prior visit of stage 1: $F_{3|1}(t) = \Pr[U_{i3}^* - T_{i3}^* \le t, X^*_{i3} = 1 | X^*_{i1} = 1]$, representing the proportion of individuals who remain in stage $3$ within time $t$ since entering stage $3$, conditional on a prior stage visit to stage $1$, and (2) cumulative incidence function from stage 3 to stage 5 conditional on the prior visit of stage 1: $P_{35|1}(t) = \Pr[U_{i3}^* - T_{i3}^* \le t, X^*_{i5} = 1, X^*_{i3} = 1 | X^*_{i1} = 1]$, representing the proportion of individuals who transition from stage $3$ to stage $5$ within time $t$ since entering stage $3$, conditional on a prior stage visit to stage $1$. Based on the property of the progressive multi-stage model and the chain rule of conditional probability, the above two quantities can be written as follows:
\begin{equation}
   \begin{aligned}
  F_{3|1}(t)  &=\Pr [U_{i3}^*-T_{i3}^*\le t| \ X^*_{i3}=1]
              \Pr [\ X^*_{i3}=1| X^*_{i1}=1]\\
              &=\Pr [U_{i3}^*-T_{i3}^*\le t| \ X^*_{i3}=1]
              \Pr [U_{i1}^*-T_{i1}^*\le \infty,\ X^*_{i3}=1| X^*_{i1}=1]\\
              &:=[1 - S_3(t)]\ P_{13}(\infty),              
   \end{aligned}\label{eq1}
\end{equation}
where $S_3(t)$ denotes the survival function for the waiting time in stage 3, and $P_{13}(t)$ represents the cumulative incidence function from stage 1 to stage 3 for the waiting time in stage 1. When $t\rightarrow \infty$, $P_{13}(\infty)$ becomes the transition probability (i.e., branching probability) from stage 1 to stage 3.
\begin{equation}
   \begin{aligned}
  P_{35|1}(t) &=\Pr [U_{i3}^*-T_{i3}^*\le t,\ X^*_{i5}=1|\ X^*_{i3}=1]
              \Pr [\ X^*_{i3}=1| X^*_{i1}=1]\\
              &:=P_{35}(t)\ P_{13}(\infty),               
   \end{aligned}\label{eq2}
\end{equation}
where $P_{35}(t)$ is the cumulative incidence function from stage 3 to stage 5 for the waiting time since entering stage 3.

In the counting process framework for analyzing multi-stage data as in the Aalen-Johansen estimator (Aalen et al., 2008), one keeps track of the number of transitions and exit times out of a state. In the present contexts, they have to be redefined in terms of stage waiting times. For uncensored data, we define them as follows:
\begin{equation}
   \begin{aligned}
    &N^*_{jj'}(t)=\sum_i I\left[U_{ij}^*-T_{ij}^*\le t,\ X^*_{ij}=1,\ X^*_{ij'}=1\right]\\
    &N^*_{j}(t)=\sum_i I\left[U_{ij}^*-T_{ij}^*\le t,\ X^*_{ij}=1\right]\\
    &Y^*_{j}(t)=\sum_i I\left[U_{ij}^*-T_{ij}^*\ge t,\ X^*_{ij}=1\right]      
   \end{aligned}.
   \label{uc: risk set}
\end{equation}

To estimate the two target quantities described in (\ref{eq1}) and (\ref{eq2}), we first derive estimators for $S_j(t)$ and $P_{jj'}(t)$. The survival function $S_j(t)$ of stage $j$ waiting times can be estimated by the Kaplan-Meier estimator
\begin{equation}
    S_j^*(t)=\prod_{s\le t}\left(1-\frac{dN_j^*(s)}{Y_j^*(s)}\right).
    \label{Kaplan-Meier estimator}
\end{equation}
The cumulative incidence function $P_{jj'}(t)$ can be estimated by the Aalen-Johansen estimator (Aalen et al., 2008) based on waiting times
\begin{equation}
    P^*_{jj'}(t)=\int_{0}^{t} S_j^*(u-) dA^*_{jj'}(u)=\int_{0}^{t} \prod_{s\le u-}\left(1-\frac{dN_j^*(s)}{Y_j^*(s)}\right)\frac{dN_{jj'}^*(u)}{Y_j^*(u)},
    \label{Aalen-Johansen estimator}
\end{equation}
where $A^*_{jj'}(t)=\int_{0}^{t}\frac{dN^*_{jj'}(t)}{Y^*_j(t)}$ is the Nelson-Aalen estimator for the cumulative transition hazard $A_{jj'}(t)$. Using (\ref{Kaplan-Meier estimator}) and (\ref{Aalen-Johansen estimator}), the estimators for $F_{3|1}(t)$ and $P_{35|1}(t)$, expressed in (\ref{eq1}) and (\ref{eq2}), for uncensored data can be written as
\begin{equation}
   \begin{aligned}
  F_{3|1}^*(t)&=\left[1- \prod_{s\le t}\left(1-\frac{dN_3^*(s)}{Y_3^*(s)}\right)\right] \left[ \int_{0}^{\infty} \prod_{s\le u-}\left(1-\frac{dN_1^*(s)}{Y_1^*(s)}\right)\frac{dN_{13}^*(u)}{Y_1^*(u)}\right],         
   \end{aligned}\label{uc: F}
\end{equation}
\begin{equation}
   \begin{aligned}
  P^*_{35|1}(t)&=\left[\int_{0}^{t} \prod_{s\le u-}\left(1-\frac{dN_3^*(s)}{Y_3^*(s)}\right)\frac{dN_{35}^*(u)}{Y_3^*(u)}\right]\ \left[\int_{0}^{\infty} \prod_{s\le u-}\left(1-\frac{dN_1^*(s)}{Y_1^*(s)}\right)\frac{dN_{13}^*(u)}{Y_1^*(u)}\right].           
   \end{aligned}\label{uc: P}
\end{equation}

\subsection{IPCW estimators for censored data}\label{method: ipcw}
For censored data, we cannot calculate $F_{3|1}^*(t)$ and $P^*_{35|1}(t)$ due to the lack of information for some $T_{ij}^{*}$, $U_{ij}^{*}$, $T_{i}^{*}$, and $X^*_{ij}$. In this section, we construct analogous estimators for $F_{3|1}(t)$ and $P_{35|1}(t)$ that are based on available data in the censored experiment. To account for the dependent censoring, we reweight the counting process and the at-risk set using the weight function $K_i(t)$ based on the inverse probability of censoring weighting (IPCW).

The weight function $K_i(t)$ represents the probability of the $i$th individual not being censored at time $t$ and is defined as: 
\begin{equation}
    K_i(t)=\prod_{s\le t}\{1-d\Lambda^c[s|{\bf Z}_i(s)]\},\nonumber 
\end{equation}
where $\Lambda^c[t|{\bf Z}_i(t)]=\int_{0}^{t} \lambda^c[u|{\bf Z}_i(u)]\text{d} u$ is the cumulative censoring hazard, $\lambda^c[u|{\bf Z}_i(u)]$ is the censoring hazard, and ${\bf Z}_i(u)$ is a vector of (predictable) covariates that may affect the hazard of being censored. In the simplest case of independent censoring, $K_i(t)$ can be estimated by the Kaplan-Meier estimator of the survival function of censoring times.

In our study, we apply Aalen's nonparametric additive hazard model (e.g., Datta and Satten, 2002; Lan and Datta, 2010; Yang et al., 2023) to model $\Lambda^c[t|{\bf Z}_i(t)]$:
\begin{equation}
    \lambda^c[t|{\bf Z}_i(t)]=\sum_{m=0}^{M}\beta_m(t)Z_{im}(t),\nonumber 
\end{equation}
where $m$ denotes the index of a covariate, and $Z_{im}(t)$ represents the $m$th covariate function for the $i$th individual ($Z_{i0}(t) = 1$). Under stage-dependent censoring, $Z_{im}(t) = I[s_i(t-) = m], 1 \leq m \leq J$, where $s_i(t)$ is the stage occupied by the $i$th individual at time $t$.

Let $B_m(t)=\int_0^t \beta_m(s)\text{d}s$. Aalen’s model estimates for ${\bf B}(t) = [B_0(t), B_1(t), \ldots, B_M(t)]$ can be expressed as:
$$\hat{\bf B}(t)=\sum_{i=1}^n I(T_i\le t) (1-\delta_i){\bf R}^{-1}(T_i){\bf Z}_i(T_i),$$
where
$${\bf R}(t)=\sum_{i=1}^n I(T_i\ge t){\bf Z}_i(t){\bf Z}^{\text T}_i(t).$$

Then we have 
\begin{equation}
\begin{aligned}
  \hat{\Lambda}^c[t|Z_i(t)]&=\sum_{m=0}^{M}\int_0^t Z_{im} (s)\text d \hat{B}_m(s),\\
              &=\sum_{i'=1}^n I(T_{i'}\le t)(1-\delta_{i'}) {\bf Z}^{\text T}_i(T_{i'}) {\bf R}^{-}(T_{i'}){\bf Z}_{i'}(T_{i'}),\ t\le T_i,          \nonumber 
\end{aligned}
\end{equation}
where ${\bf R}^{-}$ denotes the generalized inverse of $\bf R$. In our study, we use the Moore–Penrose inverse.

Therefore, the counting process and the at-risk set adjusted for IPCW in the context of dependent censoring data are given by: 
\begin{equation}
   \begin{aligned}
    &\bar{N}_{jj'}(t)=\sum_i I[U_{ij}-T_{ij}\le t,\ \delta_{ij}=1,\ \gamma_{ij'}=1]/K_i(U_{ij}-),\\
    &\bar{N}_{j}(t)=\sum_i I[U_{ij}-T_{ij}\le t,\ \delta_{ij}=1]/K_i(U_{ij}-),\\
    &\bar{Y}_{j}(t)=\sum_i I[U_{ij}-T_{ij}\ge t,\ \gamma_{ij}=1]/K_i(T_{ij}+t).   
   \end{aligned}
   \label{ipcw:risk set}
\end{equation}
After replacing $K_i(t)$ by $\hat{K}_i(t)$ in Formulas (\ref{ipcw:risk set}), we have $\hat{N}_{jj'}(t)$, $\hat{N}_{j}(t)$, and $\hat{Y}_{j}(t)$.  Substituting $\hat{N}_{jj'}$, $\hat{N}_{j}(t)$, and $\hat{Y}_{j}(t)$ for $N^*_{jj'}(t)$, $N^*_{j}(t)$, and $Y^*_{j}(t)$ respectively in (\ref{uc: F}) and (\ref{uc: P}), we derive the IPCW estimators for $F_{3|1}(t)$ and $P_{35|1}(t)$,
\begin{equation}
   \begin{aligned}
  \hat{P}_{35|1}(t)= \left[\int_{0}^{t} \prod_{s\le u-}\left(1-\frac{d\hat{N}_3(s)}{\hat{Y}_3(s)}\right)\frac{d\hat{N}_{35}(u)}{\hat{Y}_3(u)}\ \right] \left[\int_{0}^{\infty} \prod_{s\le u-}\left(1-\frac{d\hat{N}_1(s)}{\hat{Y}_1(s)}\right)\frac{d\hat{N}_{13}(u)}{\hat{Y}_1(u)} \right] ,    
   \end{aligned}
\label{IPCW: P}
\end{equation}
\begin{equation}
   \begin{aligned}
  \hat{F}_{3|1}(t)=\left[1- \prod_{s\le t}\left(1-\frac{d\hat{N}_3(s)}{\hat{Y}_3(s)}\right)\right] \left[ \int_{0}^{\infty} \prod_{s\le u-}\left(1-\frac{d\hat{N}_1(s)}{\hat{Y}_1(s)}\right)\frac{d\hat{N}_{13}(u)}{\hat{Y}_1(u)} \right].     
   \end{aligned}
   \label{IPCW: F}
\end{equation}

{\bf Lemma 1} {\it Assume that $K_i(t)$ is positive with probability 1 for each $i$ and $j$ is not a terminal stage. Then, we have
\[
\mathbb{E}[N^*_{jj'}(t)] = \mathbb{E}[\bar{N}_{jj'}(t)], \quad \mathbb{E}[N^*_{j}(t)] = \mathbb{E}[\bar{N}_{j}(t)], \quad \text{and} \quad \mathbb{E}[Y^*_{j}(t)] = \mathbb{E}[\bar{Y}_{j}(t)].
\]
}

{\bf Theorem 1} {\it Assume that $\displaystyle \sup_i\left\{\mathbb{E}[K_i^{-2}(t)]\right\}<\infty$, the assumptions in Lemma 1 hold, $j$ is not a terminal stage, and $j \neq j'$. Then, $\hat{S}_j(u)$ converges uniformly in probability to $S_j(u)$ on $[0, t]$, and $\hat{P}_{jj'}(u)$ converges uniformly in probability to $P_{jj'}(u)$ on $[0, t]$.}

To prove the consistency of $\hat{P}_{35|1}(t)$ and $\hat{F}_{3|1}(t)$ (cf., (\ref{IPCW: P}) and (\ref{IPCW: F})) for estimating $P_{35|1}(t)$ and $F_{3|1}(t)$ (cf., (\ref{eq1}) and (\ref{eq2})), we first show that the censored data counting processes and the at-risk set described in (\ref{ipcw:risk set}) are unbiased estimators of the corresponding full data counting processes and the at-risk set, as shown in Lemma 1. Using Lemma 1 and consistent estimation of $K_i(t)$, we obtain $\hat{N}_{jj'}(t)/N^*_{jj'}(t)\xrightarrow{P} 1$, $\hat{N}_{j}(t)/N^*_{j}(t)\xrightarrow{P} 1$, and $\hat{Y}_{j}(t)/Y^*_{j}(t)\xrightarrow{P} 1$. Consequently, we can establish the the consistency of $\hat{\Lambda}_j(t)$ for estimating $\Lambda_j(t)$, as well as the consistency of $\hat{A}_{jj'}(t)$ for estimating $A_{jj'}(t)$. Through the continuity result provided by the Duhamel equation (see Proposition II8.7 in Andersen et al., 1993), we obtain the consistency of $\hat{S}_j(t)$ for estimating $S_j(t)$ and of $\hat{P}_{jj'}(t)$ for estimating $P_{jj'}(t)$, described in Theorem 1. Finally, by applying the continuous mapping theorem, we can obtain the consistency of $\hat{P}_{35|1}(t)$ and $\hat{F}_{3|1}(t)$ for estimating $P_{35|1}(t)$ and $F_{3|1}(t)$. Further details of the proofs of Lemma 1 and Theorem 1 are outlined in Web Appendices A and B.

\subsection{FRE estimators for censored data}\label{method: fre}
Intuitively, individuals censored before reaching the stage of interest, such as stage $j$, could still potentially transition into stage $j$. To address this, a fractional observation can be added to the at-risk set for stage $j$ (i.e., $\bar{Y}_{j}(t)$ in (\ref{ipcw:risk set})). The idea for utilizing fractional observation is inspired by the seminal work of Datta and Satten (2000) and the subsequent research by Yang et al. (2023). In our study, we denote $\psi_{ij}$ as the fractional observation for stage $j$ from individual $i$. $\psi_{ij}$ can be interpreted as the contribution of the $i$th individual to the at-risk set for transitioning out of stage $j$. The estimation of $\psi_{ij}$ involves calculating the probability of the $i$th individual eventually reaching stage $j$ from stages preceding stage $j$. We use the abbreviated term FRE to represent the corresponding competing risk estimator, incorporating the fractional observation.

The idea of a ``unique path'' is important in constructing $\psi_{ij}$. For individuals who have been observed to reach stage $j$, $\psi_{ij}$ should be 1. Likewise, for individuals who have been observed to enter a stage which is not on the path from stage 0 to stage $j$, $\psi_{ij}$ should be 0. For a censored individual, we let $j^c$ to denote the stage where it got censored on the path from stage 0 to stage $j$ (before reaching stage $j$), and stage $j^c_{+1}$ to denote the stage immediately following stage $j^c$. For such individuals, let 
\begin{equation}
   \psi_{ij} = \left[P_{j^c j^c_{+1}}(\infty)-P_{j^c j^c_{+1}}(C_i-T_{ij^c})\right]\ \prod_{j'\in\  \mathcal{P}_{j^c_{+1}, j_{-1}}}{P_{j'j'^{+1}}(\infty)},
    \label{psi: censor}
\end{equation}
where
$$\mathcal{P}_{j^c_{+1}, j_{-1}}:= j^c_{+1}\to \ldots \to j' \to j'_{+1} \to \ldots \to j_{-1}$$
represents the unique transition path from stage $j^c_{+1}$ to stage $j_{-1}$. (\ref{psi: censor}) captures the probability of individuals censored at stage $j^c$ eventually transitioning to stage $j$ starting from their censoring time. The first factor, $\left[P_{j^c j^c_{+1}}(\infty)-P_{j^c j^c_{+1}}(C_i-T_{ij^c})\right]$, represents the probability of transitioning from stage $j^c$ to stage $j^c_{+1}$ after censoring. The second factor, $\prod_{j'\in\  \mathcal{P}_{j^c_{+1}, j_{-1}}}{P_{j'j'_{+1}}(\infty)}$, captures the probability of subsequent transitions through $\mathcal{P}_{j^c_{+1}, j_{-1}}$. This second factor is set to 1 when $j^c_{+1}$ is exactly $j$. Therefore, based on the idea of a ``unique path'', the estimation for $\psi_{ij}$ can be formulated as follows:
\begin{equation}
   \hat{\psi}_{ij}=\left\{
		\begin{aligned}
		    &\left[\hat{P}_{j^c j^c_{+1}}(\infty)-\hat{P}_{j^c j^c_{+1}}(C_i-T_{ij^c})\right]\ \prod_{j'\in\  \mathcal{P}_{j^c_{+1}, j_{-1}}}{\hat{P}_{j'j'_{+1}}(\infty)},\\
            &\ \ \ \ \text{if individual i is censored before stage $j$},\\
            &0,\ \text{if individual i is not in the path from stage 0 to stage $j$},\\
            &1,\ \text{if individual i enters stage $j$}.
		\end{aligned}
		\right.
        \label{fractional observation}
\end{equation}
For further illustration, we consider $\psi_{i3}$, the fractional observation for stage 3 from individual $i$, in the example shown in Figure \ref{fig:1-1}. If individual $i$ is not on the unique path from stage 0 to stage 3, denoted $\mathcal{P}_{0,3}$ (for example, if they transition to stage 2 or stage 4), then $\psi_{i3} = 0$. If individual $i$ is observed to enter stage 3, $\psi_{i3} = 1$. If individual $i$ is censored at stage 0, $\psi_{i3}=\left[P_{01}(\infty) - P_{01}(C_i-T_{i0})\right] P_{13}(\infty)$. Similarly, if individual $i$ is censored at stage 1, $\psi_{i3}=P_{13}(\infty) - P_{13}(C_i-T_{i1})$.

After incorporating the fractional observation $\psi_{ij}$, the at-risk set adjusted for FRE, denoted as $\bar{Y}^{F}_{j}(t)$, can be estimated by
\begin{equation}
\hat{Y}^{F}_{j}(t)=\sum_i \hat{\psi}_{ij}\Biggl\{\frac{I[U_{ij}-T_{ij}\ge t,\ \gamma_{ij}=1]}{\hat{K}_i(T_{ij}+t)}+ I(\gamma_{ij^c}=1, \delta_{ij^c}=0)\hat{S}_j(t) \Biggl\},   
\label{fre: risk set}
\end{equation}
in which $\frac{I[U_{ij}-T_{ij}\ge t,\ \gamma_{ij}=1]}{\hat{K}_i(T_{ij}+t)}$ represents the reweighted at-risk set for individuals observed to enter stage $j$, and $I(\gamma_{ij^c}=1, \delta_{ij^c}=0)$ is the indicator function, which equals to 1 if individual $i$ is censored in stage $j^c$ and 0 otherwise. It should be noted that for individuals censored in stage $j^c$, they still have the probability of remaining in stage $j$ for longer than the waiting time $t$. 
An implicit assumption for the validity of (\ref{fre: risk set}) is that future entry/exit times (i.e., the waiting times at stage $j$) are independent of the censoring, given the covariate information up to that time point. Hence, consistent with the definition of the at-risk set, we include the multiplication of $\hat{S}_j(t)$. 
$\mathbb{E}[Y^*_{j}(t)]=\mathbb{E}[\bar{Y}^{F}_{j}(t)]$ is also possibly demonstrated by a similar approach as in the proof for IPCW estimators in Section \ref{method: ipcw}. Thus, we obtain the FRE estimators, denoted as $\hat{F}^F_{3|1}(t)$ and $\hat{P}^F_{35|1}(t)$, by replacing $\hat{Y}_{j}(t)$ with $\hat{Y}^{F}_{j}(t)$ in the formulas for $\hat{F}_{3|1}(t)$ and $\hat{P}_{35|1}(t)$ (i.e., (\ref{IPCW: P}) and (\ref{IPCW: F})).

\subsection{Extension to a general multi-stage model}
In the above sections, we use an illustrative six-stage survival model to derive empirical estimators for uncensored data (i.e., Section \ref{method: def}), IPCW estimators for censored data (i.e., Section \ref{method: ipcw}), and FRE estimators for censored data (i.e., Section \ref{method: fre}) regarding the stage waiting time distribution conditional on the prior visit, $F_{3|1}(t)$, and the cumulative incidence functions conditional on the prior visit, $P_{35|1}(t)$. These estimators can be easily extended to the general multi-stage model under a progressive structure. Let $j_{+1}$ and $j_{-1}$ denote the stages immediately following and preceding stage $j$, respectively. Stage $k$ represents a stage of interest prior to stage $j$. Based on the chain rule of conditional probability and the property of the progressive tree structure, where each stage lies in a unique path originating from the root node, we have
\begin{equation}
   \begin{aligned}
 P_{jj_{+1}|k}(t)=P_{jj_{+1}}(t)\ \prod_{j'\in \mathcal{P}_{k,j_{-1}}}{P_{j'j'_{+1}}(\infty)},   
   \end{aligned}
\nonumber
\end{equation}

\begin{equation}
   \begin{aligned}
 F_{j|k}(t)=\left[1-S_{j}(t)\right]\ \prod_{j'\in \mathcal{P}_{k,j_{-1}}}{P_{j'j'_{+1}}(\infty)},     
   \end{aligned}
\nonumber
\end{equation}
where
$$\mathcal{P}_{k,j_{-1}}:= k\to \ldots \to j' \to j'_{+1} \to \ldots \to j_{-1}.$$

For the IPCW estimation of $P_{jj_{+1}|k}(t)$ and $F_{j|k}(t)$ for censored data, we have $\hat{P}_{jj_{+1}}(t)=\int_{0}^{t} \prod_{s\le u-}\left(1-\frac{d\hat{N}_j(s)}{\hat{Y}_j(s)}\right)\frac{d\hat{N}_{jj_{+1}}(u)}{\hat{Y}_j(u)}$, and $\hat{S}_{j}(t)= \prod_{s\le t}\left(1-\frac{d\hat{N}_j(s)}{\hat{Y}_j(s)}\right)$ based on the reweighted counting process and the at-risk set given in (\ref{ipcw:risk set}). Using these expressions, the IPCW estimators $\hat{P}_{jj_{+1}|k}(t)$ and $\hat{F}_{j|k}(t)$ can be derived accordingly. Similarly, the empirical estimators for uncensored data and FRE estimators for censored data can be analogously generated using (\ref{uc: risk set}) and (\ref{fre: risk set}).

\section{Simulation Studies}\label{simulation}
The numerical performance of the proposed methods described in Subsections \ref{method: ipcw} and \ref{method: fre} is evaluated using two different metrics. The simulation setups are outlined in Subsection \ref{sim: design}, and the corresponding results are presented in Subsection \ref{sim: est}. Additional simulation results are provided in Web Appendix C.

\subsection{Simulation setups}\label{sim: design}
To validate our methods, we conduct simulation studies across various settings and scenarios. The performances of our methods are assessed using data generated from the aforementioned illustrative example of the six-stage progressive model in Figure \ref{fig:1-1}. The simulation designs are presented below.

{\bf Waiting time:}  We generate the waiting time $w_{ij}=U_{ij}^*-T_{ij}^*$ for $i$th individual in stage $j$ under two different models: the Markov model and the semi-Markov model. To incorporate individual-level heterogeneity, we introduce a frailty parameter $z_i$. Subsequently, $w_{ij}$ is multiplied by $z_i$ to get the final individual stage waiting time, which is $z_i w_{ij}$. $z_i$ is assumed to follow a log-normal distribution with a log-mean of 0 and a log-scale of $\tau$. We consider two values of $\tau$, $\tau = 0$ and $\tau = 1$. The main paper presents the full results for $\tau = 0$, where $ z_i = 1$. Note that stages $j = 0, 1, 3$ are considered, as stages 2, 4, and 5 are terminal, and the time of leaving stage $j$ (i.e., $U_{ij}^*$) is equal to the time of entering the next stage (i.e., $T_{ij_{+1}}^*$). Under the Markov model, event times $U_{ij}^*$ are generated from a Weibull distribution with a shape parameter of 2 and a scale parameter of 4, denoted as WB(2, 4), or from a log-normal distribution with a log-mean parameter of 0.9 and a log-scale parameter of 0.5, denoted as LN(0.9, 0.5). To ensure the Markov property in these multi-stage models, we can apply the transformation (as discussed in Mostajabi and Datta, 2013; Anyaso-Samuel et al., 2023):
\begin{equation}
    U_{ij_{+1}}^*=D^{-1}\Bigl\{D(U_{ij}^*)+R\left[0, 1-D(U_{ij}^*)\right]\Bigl\},
    \label{markov}
\end{equation}
where $D$ denotes the cumulative distribution function for $U_{ij}^*$, $D^{-1}$ is the corresponding quantile function, and $R$ is a random number generated from the uniform distribution $\left[0, 1-D(U_{ij}^*)\right]$. For the semi-Markov model, waiting times $w_{i0}$, $w_{i1}$, and $w_{i3}$ are generated independently from WB(2,4), WB(3, 4), and WB(1, 2) or from LN(0.9, 0.5), LN(0.8, 0.5), and LN(0.7, 0.5). 

{\bf Branching probability:} In these setups, every transient stage allows for two possible branches. We control the branching by a Bernoulli variable, denoted as $B_{i,jj_{+1}}$. Specifically, we assume $B_{i,jj_{+1}} \sim \text{Bernoulli}(\Phi_{i,\ jj_{+1}})$, where $\Phi_{i,\ jj_{+1}}$ represents the branching probability from stage $j$ to $j_{+1}$ for individual $i$. If $B_{i,01}=1$, the individual enters stage 1 from stage 0; likewise, if $B_{i,01}=0$, the individual enters stage 2 from stage 0. The same principle applies for $j=1, 3$. We model $\Phi_{i,\ jj_{+1}}$ using a logit function, $\text{logit}(\Phi_{i,\ jj_{+1}}) = \alpha + \beta\ w_{ij}$, where $w_{ij}$ is the waiting time for individual $i$ in stage $j$. In the main paper, we present complete results for the simulation cases where  $\alpha$ and $\beta$ equals to 0, which means that $B_{i,jj_{+1}} \sim \text{Bernoulli}(0.5)$, where all branching probabilities are set to 0.5.

{\bf Censoring time:} For simplicity in bookkeeping, we let the censoring time follow the waiting time distribution. Therefore, censoring times are generated from a Weibull or log-normal distribution with different parameters to control the censoring percentage. Independent censoring and stage-dependent censoring are the two types considered. For each type, we explore two censoring schemes: low and high censoring rates. The distributions and parameters used to generate censoring times are provided in Table \ref{censor}. Under independent censoring, censoring times are independent of the multi-stage waiting times. Under stage-dependent censoring simulation, we first simulate the censoring time at stage 0. If $C_i < U_{i0}^*$, individuals are censored at stage 0; otherwise, they are observed to leave stage 0 before censoring occurs. We then simulate the censoring times at stages 1 and 3 using an approach similar to Formula (\ref{markov}), where the quantile function $D$ is chosen according to Table \ref{censor}.

\begin{table}[htb]
\centering
\setlength{\belowcaptionskip}{+0.2cm}
\caption{Parameters of the Weibull and log-normal distributions for generating censoring times.}
\begin{tabular}{ccccc}
\hline
& & &\multicolumn{2}{c}{Censoring distribution} \\
\cline{4-5}
Stage  & Censoring type & Censoring scheme & WB & LN \\
\hline
            & Independent & High & WB(3,6) & LN(1,0.8)   \\
Stage 0     & Independent & Low  & WB(3,9) & LN(1.8,1)   \\
            & Dependent   & High & WB(3,5) & LN(1,0.6)   \\
            & Dependent   & Low  & WB(3,7) & LN(1.8,0.8) \\
\hline
            & Independent & High & WB(3,6) & LN(1,0.8)   \\
Stage 1     & Independent & Low  & WB(3,9) & LN(1.8,1)   \\
            & Dependent   & High & WB(2,3) & LN(0.9,0.5) \\
            & Dependent   & Low  & WB(2,5) & LN(1.2,0.6) \\
\hline
            & Independent & High & WB(3,6) & LN(1,0.8)   \\
Stage 3     & Independent & Low  & WB(3,9) & LN(1.8,1)   \\
            & Dependent   & High & WB(2,2) & LN(0.8,0.4) \\
            & Dependent   & Low  & WB(2,3) & LN(0.6,0.4) \\
\hline
\end{tabular}
\label{censor}
\end{table}

\subsection{Results about $\hat{F}_{3|1}(t)$ and $\hat{P}_{35|1}(t)$}\label{sim: est}

Tables \ref{L1:markov:F} -- \ref{L1:semi:P} present the $L_1$ norm of the estimation errors corresponding to the two estimators, denoted by $\Delta$, for various simulation settings and sample sizes ($n=100, 300, 600$). $\Delta$ serves as an overall performance measure to assess the two different estimators. We evaluate $\Delta$ at the 10th, 20th, \ldots, and 90th percentiles, $t_k$, of the empirical distribution function $F_{3|1}(t)$ (or $P_{35|1}(t)$) with a large sample size of 10,000. $\Delta$ is computed based on 5,000 Monte Carlo iterations as follows: 
\begin{equation}
  \Delta =\frac{1}{9 \times 5,000}\sum_{i=1}^{5,000}\sum_{k=1}^{9}\left| \theta (t_k ) -\hat{\theta} (t_k )\right|,  \nonumber 
\end{equation}
where $\theta$ represents $F_{3|1}(t)$ (or $P_{35|1}(t)$), and $\hat{\theta}$ denotes either the IPCW estimator or the FRE estimator of $F_{3|1}(t)$ (or $P_{35|1}(t)$). Column 4 in Tables \ref{L1:markov:F} -- \ref{L1:semi:P} summarizes the censoring rates in each simulation setting. The censoring rates range from 23.6\% to 35.4\% under the low censoring scheme and from 46.2\% to 63\% under the high censoring scheme. From Tables \ref{L1:markov:F} -- \ref{L1:semi:P}, for both Markov and semi-Markov models, $\Delta$ decreases as the sample size increases for IPCW and FRE estimators. The IPCW estimators perform better than the FRE estimators with smaller values of $\Delta$. One possible explanation is that the calculation of the fractional observation for the FRE estimators relies on the idea of a ``unique path'' and depends on estimates from the previous stages. Thus, the estimation errors in the fractional observation may propagate along the paths. 

\begin{figure}
    \centering
    \begin{subfigure}[t]{0.45\textwidth}
        \centering
        \includegraphics[scale=.55]{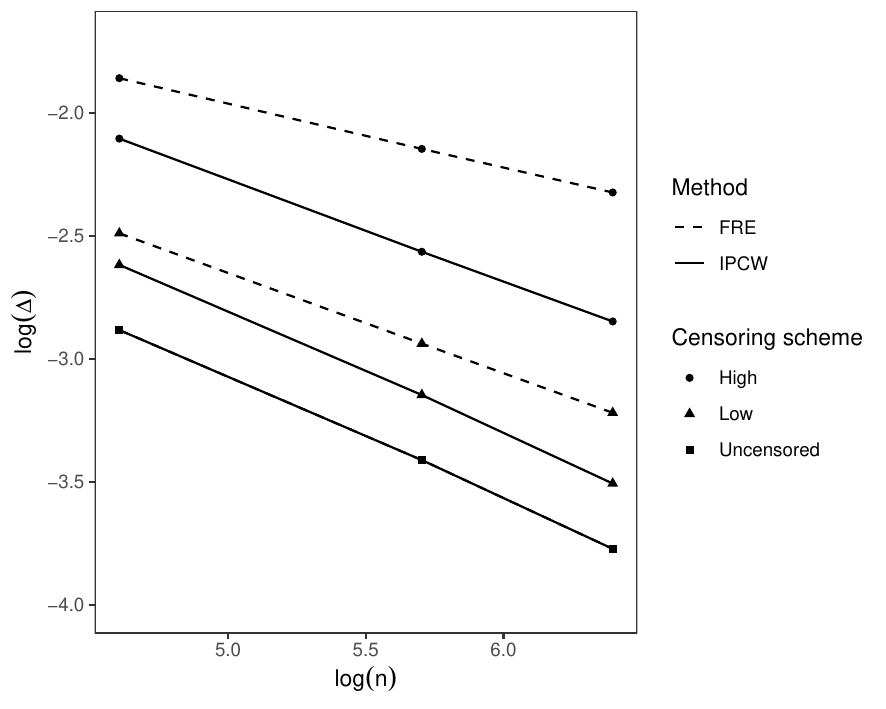}
        \caption{WB waiting and censoring times; independent censoring} \label{fig:L1:1}
    \end{subfigure}
    \hfill
    \begin{subfigure}[t]{0.45\textwidth}
        \centering
        \includegraphics[scale=.55]{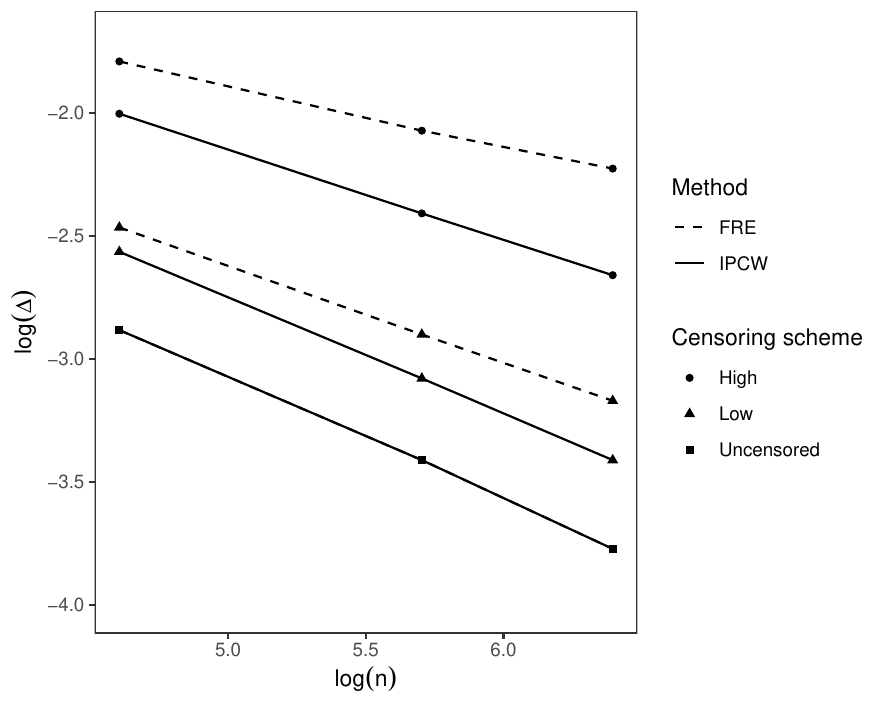} 
        \caption{WB waiting and censoring times; stage-dependent censoring} \label{fig:L1:2}
    \end{subfigure}

    \vspace{1cm}
    \begin{subfigure}[t]{0.45\textwidth}
        \centering
        \includegraphics[scale=.55]{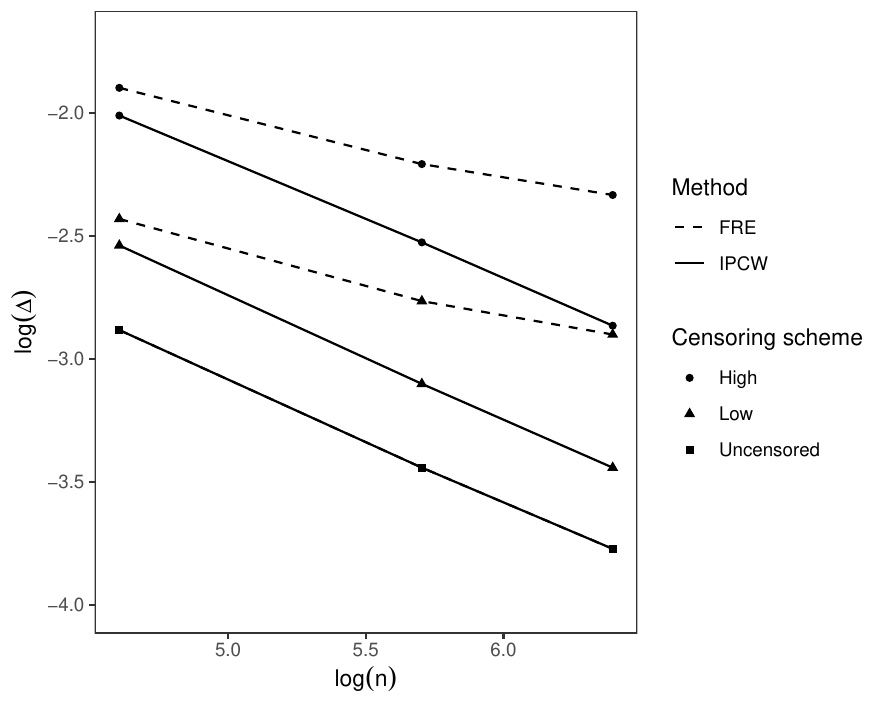}
        \caption{LN waiting and censoring times; independent censoring} \label{fig:L1:3}
    \end{subfigure}
    \hfill
    \begin{subfigure}[t]{0.45\textwidth}
        \centering
        \includegraphics[scale=.55]{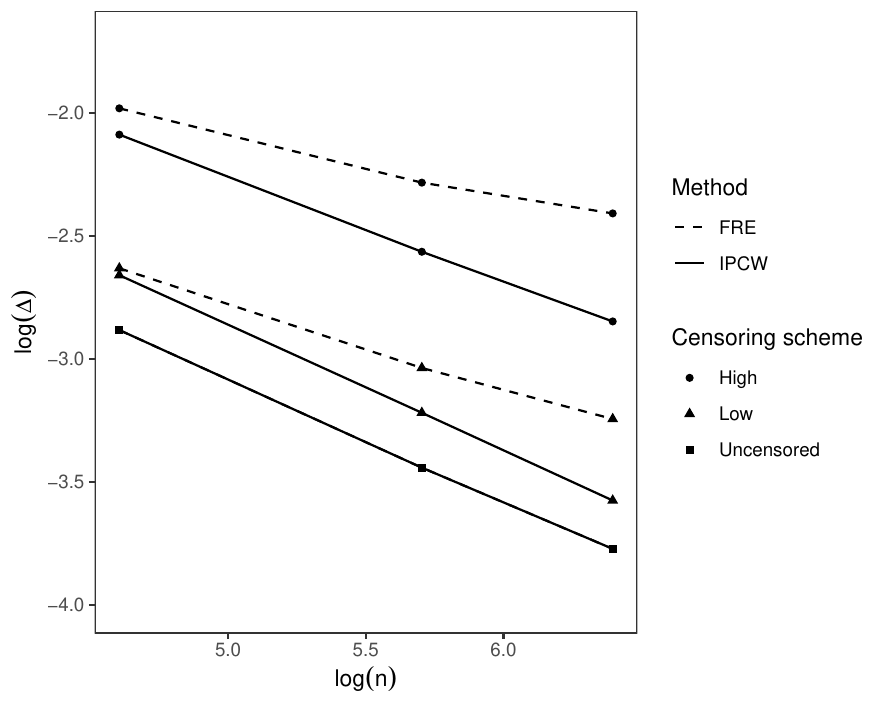}
        \caption{LN waiting and censoring times; stage-dependent censoring} \label{fig:L1:4}
    \end{subfigure}
    \caption{Relationships between the logarithms of the $L_1$ norm of the estimation errors and the logarithms of the sample size for the IPCW and FRE estimators of $F_{3|1}(t)$ under the Markov model}
    \label{fig:L1}
\end{figure}

Figure~\ref{fig:L1} shows approximately linear relationships between the logarithms of the $L_1$ norm of the estimation errors and the logarithms of the sample size $n$ for each estimator of $F_{3|1}(t)$ in different simulation settings for the Markov model (see Table~\ref{L1:markov:F} for the corresponding $\Delta$ values), indicating that the $L_1$ norm of the estimation errors converges to zero at a rate of $n^{-b}$ for some constant $b$. Due to page limitations, similar patterns under other simulation settings are presented in Web Figures 1 -- 3 of Appendix C, with detailed $\Delta$ values reported in Tables~\ref{L1:markov:P} -- \ref{L1:semi:P}. 

It should also be noted that the performance for both IPCW and FRE estimators improves with lower censoring rates, since a lower censoring rate provides more complete data and reduces potential bias and uncertainty in the estimators. The main paper presents the complete simulation results in the general setting with $\alpha = \beta = \tau = 0$, which corresponds to equal branching probabilities at different stages (i.e, 0.5) and no frailties in stage waiting times. Additional simulations under more complex scenarios (e.g., $\alpha = \beta = 1$, $\tau = 1$) are presented in Web Figures 4 and 5 of Web Appendix C, where we also observe similar linear decreasing trends between the logarithms of the $L_1$ norm of the estimation errors and the logarithms of the sample sizes.

\begin{table}[htb]
\scriptsize
\centering
\setlength{\belowcaptionskip}{+0.2cm}
\caption{$L_1$ norm of the estimation errors for the IPCW and FRE estimators of $F_{3|1}(t)$ under the Markov model}
\begin{tabular}{cccccccccc}
\hline
Waiting/Censoring time &\multicolumn{3}{c}{Censoring scenario}&\multicolumn{2}{c}{$n=100$}&\multicolumn{2}{c}{$n=300$}&\multicolumn{2}{c}{$n=600$} \\

Distribution & Type & Scheme & Censoring rate & IPCW & FRE & IPCW & FRE & IPCW & FRE\\
\hline
WB & Independent            & Low & 0.241 & 0.073 & 0.083 & 0.043 & 0.053 & 0.030 & 0.040 \\
WB & Independent            & High& 0.481 & 0.122 & 0.156 & 0.077 & 0.117 & 0.058 & 0.098 \\
WB & Stage-dependent        & Low & 0.276 & 0.077 & 0.085 & 0.046 & 0.055 & 0.033 & 0.042 \\
WB & Stage-dependent        & High& 0.533 & 0.135 & 0.167 & 0.090 & 0.126 & 0.070 & 0.108 \\
WB &\multicolumn{2}{c}{Uncensored}& 0.000 & 0.056 & 0.056 & 0.033 & 0.033 & 0.023 & 0.023 \\
\hline
LN & Independent            & Low & 0.339 & 0.079 & 0.088 & 0.045 & 0.063 & 0.032 & 0.055 \\
LN & Independent            & High& 0.620 & 0.134 & 0.150 & 0.080 & 0.110 & 0.057 & 0.097 \\
LN & Independent            & Low & 0.236 & 0.070 & 0.072 & 0.040 & 0.048 & 0.028 & 0.039 \\
LN & Independent            & High& 0.570 & 0.124 & 0.138 & 0.077 & 0.102 & 0.058 & 0.090 \\
LN &\multicolumn{2}{c}{Uncensored}& 0.000 & 0.056 & 0.056 & 0.032 & 0.032 & 0.023 & 0.023 \\
\hline
\end{tabular}
\label{L1:markov:F}
\end{table}

\begin{table}[htb]
\scriptsize
\centering
\setlength{\belowcaptionskip}{+0.2cm}
\caption{$L_1$ norm of the estimation errors of the IPCW and FRE estimators of $P_{35|1}(t)$ under the Markov model.}
\begin{tabular}{cccccccccccccc}
\hline
Waiting/Censoring time &\multicolumn{3}{c}{Censoring scenario}&\multicolumn{2}{c}{$n=100$}&\multicolumn{2}{c}{$n=300$}&\multicolumn{2}{c}{$n=600$} \\
Distribution & Type & Scheme & Censoring rate & IPCW & FRE & IPCW & FRE & IPCW & FRE\\
\hline
WB & Independent            & Low & 0.241 & 0.068 & 0.075 & 0.040 & 0.048 & 0.028 & 0.037 \\
WB & Independent            & High& 0.481 & 0.119 & 0.141 & 0.078 & 0.107 & 0.059 & 0.090 \\
WB & Stage-dependent        & Low & 0.276 & 0.070 & 0.076 & 0.042 & 0.049 & 0.031 & 0.038 \\
WB & Stage-dependent        & High& 0.533 & 0.126 & 0.149 & 0.086 & 0.114 & 0.069 & 0.099 \\
WB &\multicolumn{2}{c}{Uncensored}& 0.000 & 0.048 & 0.048 & 0.029 & 0.029 & 0.021 & 0.021 \\
\hline
LN & Independent            & Low & 0.339 & 0.067 & 0.078 & 0.039 & 0.056 & 0.028 & 0.049 \\
LN & Independent            & High& 0.620 & 0.115 & 0.130 & 0.067 & 0.110 & 0.047 & 0.097 \\
LN & Stage-dependent        & Low & 0.236 & 0.058 & 0.064 & 0.034 & 0.042 & 0.024 & 0.035 \\
LN & Stage-dependent        & High& 0.570 & 0.102 & 0.119 & 0.064 & 0.088 & 0.048 & 0.078 \\
LN &\multicolumn{2}{c}{Uncensored}& 0.000 & 0.047 & 0.047 & 0.028 & 0.028 & 0.020 & 0.020 \\
\hline
\end{tabular}
\label{L1:markov:P}
\end{table}

\begin{table}[htb]
\scriptsize
\centering
\setlength{\belowcaptionskip}{+0.2cm}
\caption{$L_1$ norm of the estimation errors of the IPCW and FRE estimators of $F_{3|1}(t)$ under the semi-Markov model.}
\begin{tabular}{cccccccccccccc}
\hline
Waiting/Censoring time &\multicolumn{3}{c}{Censoring scenario}&\multicolumn{2}{c}{$n=100$}&\multicolumn{2}{c}{$n=300$}&\multicolumn{2}{c}{$n=600$} \\
Distribution & Type & Scheme & Censoring rate & IPCW & FRE & IPCW & FRE & IPCW & FRE\\
\hline
WB & Independent            & Low & 0.246 & 0.078 & 0.086 & 0.047 & 0.058 & 0.033 & 0.048 \\
WB & Independent            & High& 0.462 & 0.128 & 0.147 & 0.086 & 0.112 & 0.067 & 0.098 \\
WB & Stage-dependent        & Low & 0.282 & 0.080 & 0.087 & 0.049 & 0.059 & 0.036 & 0.049 \\
WB & Stage-dependent        & High& 0.509 & 0.133 & 0.151 & 0.092 & 0.118 & 0.076 & 0.105 \\
WB &\multicolumn{2}{c}{Uncensored}& 0.000 & 0.055 & 0.055 & 0.032 & 0.032 & 0.023 & 0.023 \\
\hline
LN & Independent            & Low & 0.354 & 0.075 & 0.081 & 0.042 & 0.057 & 0.030 & 0.050 \\
LN & Independent            & High& 0.630 & 0.124 & 0.127 & 0.074 & 0.092 & 0.053 & 0.079 \\
LN & Stage-dependent        & Low & 0.247 & 0.066 & 0.068 & 0.038 & 0.045 & 0.027 & 0.038 \\
LN & Stage-dependent        & High& 0.584 & 0.117 & 0.120 & 0.074 & 0.089 & 0.054 & 0.076 \\
LN &\multicolumn{2}{c}{Uncensored}& 0.000 & 0.054 & 0.054 & 0.031 & 0.031 & 0.022 & 0.022 \\
\hline
\end{tabular}
\label{L1:semi:F}
\end{table}

\begin{table}[htb]
\scriptsize
\centering
\setlength{\belowcaptionskip}{+0.2cm}
\caption{$L_1$ norm of the estimation errors of the IPCW and FRE estimators of $P_{35|1}(t)$ under the semi-Markov model.}
\begin{tabular}{cccccccccccccc}
\hline
Waiting/Censoring time &\multicolumn{3}{c}{Censoring scenario}&\multicolumn{2}{c}{$n=100$}&\multicolumn{2}{c}{$n=300$}&\multicolumn{2}{c}{$n=600$} \\
Distribution & Type & Scheme & Censoring rate     & IPCW  & FRE    & IPCW  & FRE   & IPCW  & FRE \\
\hline
WB & Independent            & Low & 0.246 & 0.069 & 0.077  & 0.041 & 0.053 & 0.030 & 0.043\\
WB & Independent            & High& 0.462 & 0.113 & 0.132  & 0.076 & 0.103 & 0.059 & 0.090\\
WB & Stage-dependent        & Low & 0.282 & 0.069 & 0.078  & 0.043 & 0.053 & 0.032 & 0.044\\
WB & Stage-dependent        & High& 0.509 & 0.115 & 0.135  & 0.079 & 0.107 & 0.065 & 0.095\\
WB &\multicolumn{2}{c}{Uncensored}& 0.000 & 0.046 & 0.046  & 0.027 & 0.027 & 0.019 & 0.019 \\
\hline
LN & Independent            & Low & 0.354 & 0.064 & 0.071  & 0.037 & 0.051 & 0.026 & 0.045\\
LN & Independent            & High& 0.630 & 0.105 & 0.107  & 0.063 & 0.077 & 0.045 & 0.067\\
LN & Stage-dependent        & Low & 0.247 & 0.056 & 0.059  & 0.033 & 0.040 & 0.023 & 0.034\\
LN & Stage-dependent        & High& 0.584 & 0.095 & 0.101  & 0.061 & 0.075 & 0.045 & 0.065\\
LN &\multicolumn{2}{c}{Uncensored}& 0.000 & 0.044 & 0.044  & 0.026 & 0.026 & 0.019 & 0.019\\
\hline
\end{tabular}
\label{L1:semi:P}
\end{table}

Web Figures 6 -- 13 in the Web Appendix C further show the efficacy of the IPCW estimator and FRE estimator of $F_{3|1}(t)$ (or $P_{35|1}(t)$). In these figures, the solid curve represents the empirically true distribution of $F_{3|1}(t)$ (or $P_{35|1}(t)$), obtained from uncensored data simulations with a large sample size of 10,000 individuals. The dashed and dotted curves depict the IPCW and FRE estimates of $F_{3|1}(t)$ (or $P_{35|1}(t)$), respectively, obtained from the observed data with the sample size of 2,000. The abbreviations in the top-left corner of each panel denote the simulation design. For example, ``WB.WB.DST.H'' (or ``LN.LN.I.L'') represents the scenario where waiting times follow a Weibull (or log-normal) distribution and censoring times also follow a Weibull (or log-normal) distribution under the dependent (or independent) censoring type with a high (or low) censoring rate scheme. The figures clearly show that the IPCW estimates align more closely with the true distribution compared to the FRE estimates under different simulation settings. Moreover, the performance of both estimators tends to be better when the censoring rate is lower.

\section{Case Studies}\label{case studies}
In this section, we apply our proposed estimation methods discussed in Section \ref{methods} to a real data example that involves multiple progressive survival stages. An additional case study about the breast cancer data is provided in Appendix D of Supplementary Material to further illustrate the methods.

\begin{figure}[htb]
\centering
\includegraphics[scale=.50]{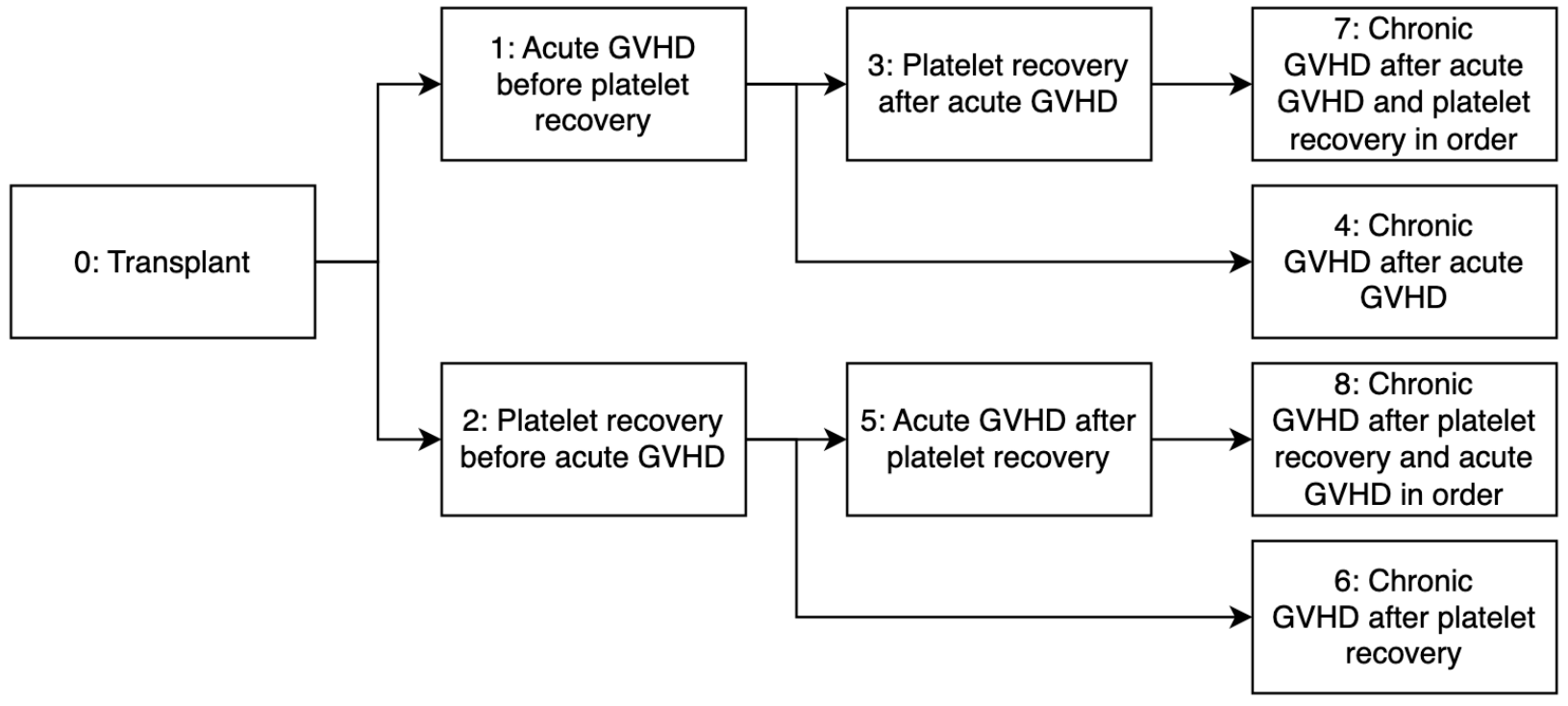}
\caption{Diagram of the nine-stage survival model for the transplant graft-versus-host disease data.}
\label{bmt9stage}
\end{figure}

The real data example involves 136 cancer patients who received bone marrow transplants, as detailed in Klein and Moeschberger (2003). The dataset is available in the \textit{R}-package \textbf{KMsurv}. Following Datta and Satten (2002), we construct a nine-stage progressive survival model based on the following three post-transplantation events: platelets recovery, the onset of acute graft-versus-host disease (GVHD), and the onset of chronic GVHD. The diagram of stages and transitions in the transplant graft-versus-host disease is presented in Figure \ref{bmt9stage}. Transition times in the original dataset were right-censored, with follow-up times ranging from 1 to 2640 days. All patients start from stage 0, representing the stage of no post-transplantation events. Stage 1 is entered when acute GVHD develops before platelet recovery. Stage 2 is entered if the patient’s platelet recovers before acute GVHD develops. After stage 1, stage 3 is for patients whose platelet recovers after acute GVHD develops, while stage 4 is for patients whose chronic GVHD develops after acute GVHD develops. After stage 2, stage 5 is for patients whose acute GVHD develops after platelet recovery, while stage 6 is for patients whose chronic GVHD develops after platelet recovery. Stage 7 is for patients whose chronic GVHD develops after acute GVHD develops (i.e., stage 1) and platelet recovery (i.e., stage 3) in order. Stage 8 is for patients whose chronic GVHD develops after platelet recovery (i.e., stage 2) and acute GVHD (i.e., stage 5) in order. Patients do not necessarily progress to stage 7 or 8 as they may remain in any stage for any amount of time. Those patients who died or experienced relapse are considered censored in this example.

Table \ref{bmt: transition} summarizes the observed numbers of transitions between stages in the transplant graft-versus-host disease data, where the diagonal elements represent the number of censored observations within each stage, and the off-diagonal elements are the numbers of observed transitions between stages. No transitions are observed from stages 4, 6, 7, and 8, as they are defined as terminal stages in the nine-stage survival model (i.e., Figure \ref{bmt9stage}). Only 7 patients developed acute GVHD after their transplantation (moved from stage 0 to stage 1), while 116 patients' platelets recovered after their transplantation (moved from stage 0 to stage 2). Therefore, we focus on the transitions after stage 2. Specifically, we are interested in the following quantities: $F_{2|0}$, $F_{5|2}$, $P_{25|0}$, $P_{26|0}$, and $P_{58|2}$. 

\begin{table}[htb]
\centering
\setlength{\belowcaptionskip}{+0.2cm}
\caption{Observed transitions in the transplant graft-versus-host disease data.}
\begin{tabular}{c ccc ccc ccc}
\hline
& \multicolumn{9}{c}{To} \\
\cline{2-10}
From & 0    & 1    & 2    & 3    & 4    & 5    & 6    & 7    & 8    \\
\hline
0    & 13   & 7    & 116  & 0    & 0    & 0    & 0    & 0    & 0    \\
1    & 0    & 2    & 0    & 3    & 2    & 0    & 0    & 0    & 0    \\
2    & 0    & 0    & 54   & 0    & 0    & 19   & 44   & 0    & 0    \\
3    & 0    & 0    & 0    & 2    & 0    & 0    & 0    & 1    & 0    \\
4    & 0    & 0    & 0    & 0    & 0    & 0    & 0    & 0    & 0    \\
5    & 0    & 0    & 0    & 0    & 0    & 8    & 0    & 0    & 11    \\
6    & 0    & 0    & 0    & 0    & 0    & 0    & 0    & 0    & 0    \\
7    & 0    & 0    & 0    & 0    & 0    & 0    & 0    & 0    & 0    \\
8    & 0    & 0    & 0    & 0    & 0    & 0    & 0    & 0    & 0    \\
 \hline
\end{tabular}
\label{bmt: transition}
\end{table}

\begin{figure}[htb]
\centering
\includegraphics[scale=.58]{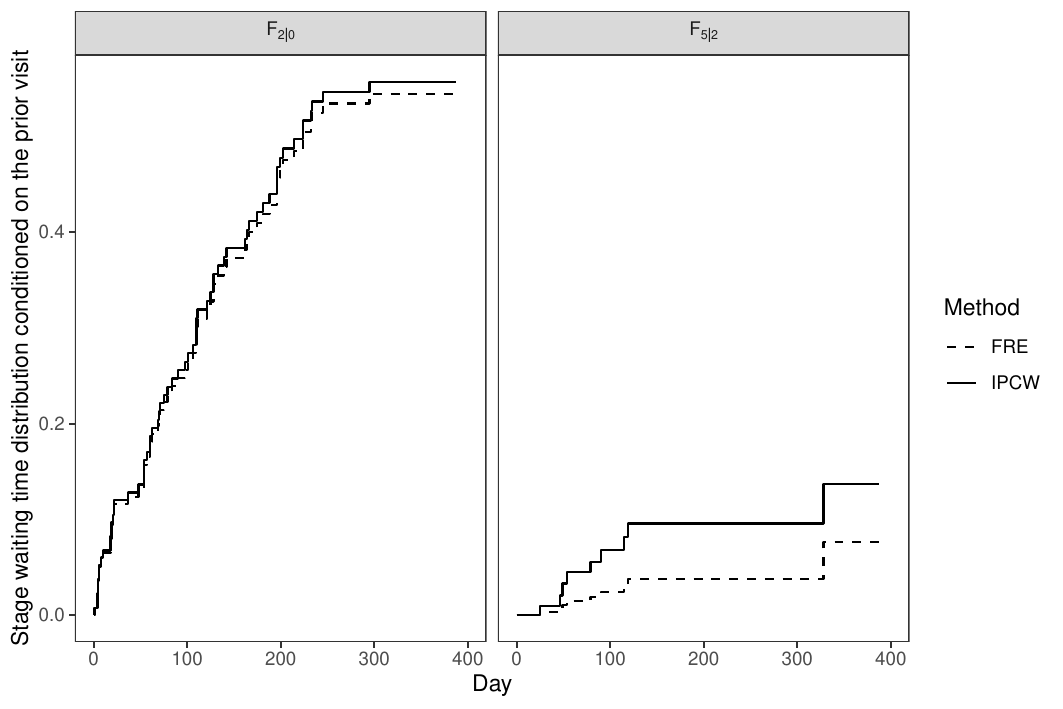} 
\caption{IPCW estimates of the stage waiting time distributions conditional on the prior stage visit mentioned in Section \ref{case studies}.}
\label{bmt_f}
\end{figure}

\begin{figure}[htb]
\centering
\includegraphics[scale=.58]{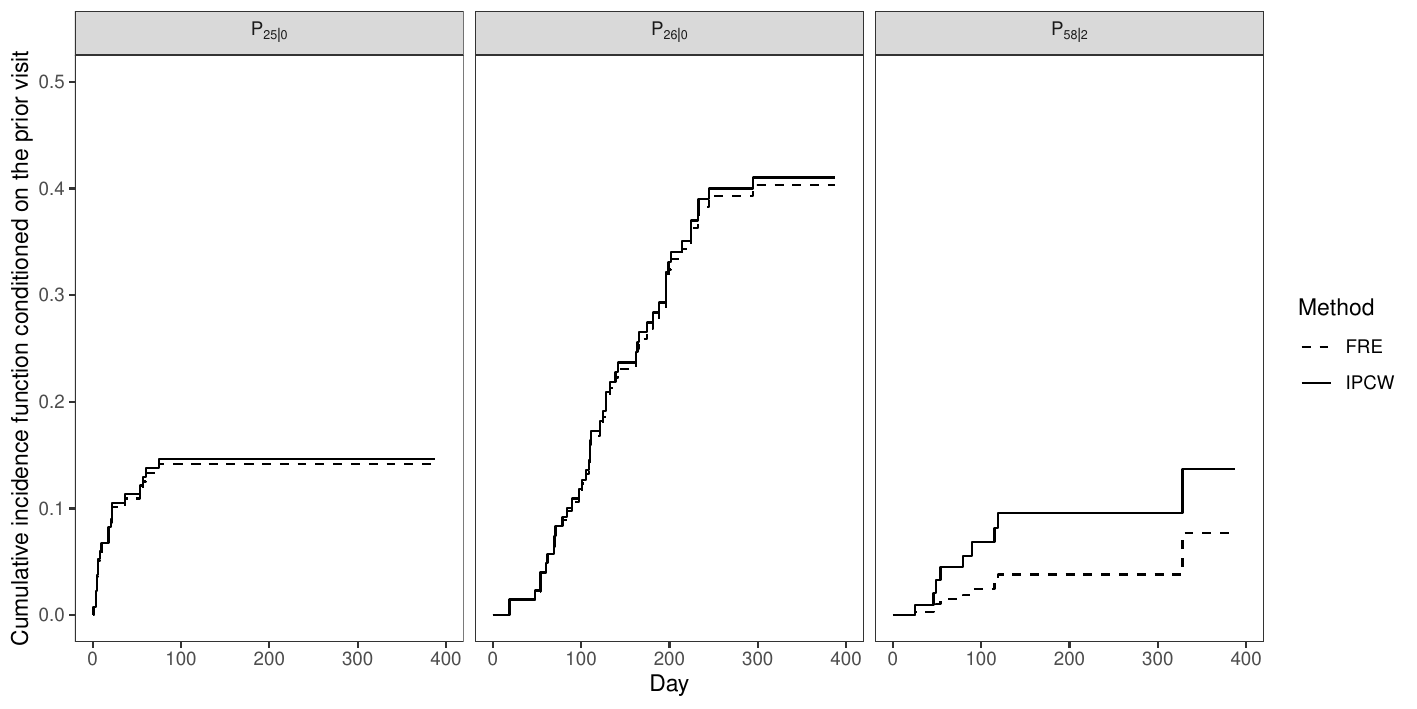}
\caption{IPCW estimates of the cumulative incidence functions conditional on the prior stage visit mentioned in Section \ref{case studies}.}
\label{bmt_p}
\end{figure}

Here are several interesting findings from the results shown in Figures \ref{bmt_f} and \ref{bmt_p}, which display the IPCW and FRE estimation results for the aforementioned stage waiting time distributions and cumulative incidence functions conditional on prior stage visits. We first focus on interpreting the IPCW estimation results, represented by the solid lines in Figures \ref{bmt_f} and \ref{bmt_p}. From Figure \ref{bmt_f}, we observe that for $\hat{F}_{2|0}$, there is no jump around the 300th waiting day since entering stage 2, and after this point, $\hat{F}_{2|0}$ remains constant at approximately 0.55. This indicates that around 55\% of post-transplant (stage 0) patients will eventually develop platelet recovery before acute GVHD (stage 2). Similarly, for $\hat{F}_{5|2}$, there is no jump around the 330th waiting day since entering stage 5, showing that around 14\% of patients who first experience platelet recovery (stage 2) will eventually develop acute GVHD (stage 5). From the comparison of $P_{25|0}$ and $P_{26|0}$ among post-transplant patients (stage 0) in Figure \ref{bmt_p}, we observe that before around the 110th waiting day following the onset of platelet recovery before acute GVHD (stage 2), patients are more likely to transition to acute GVHD (stage 5) than to chronic GVHD (stage 6). However, after around the 110th waiting day, the probability of transitioning to chronic GVHD (stage 6) becomes higher than that of transitioning to acute GVHD (stage 5). For those interested in patients after transplantation (stage 0) who eventually develop chronic GVHD (stages 6 or 8), we can compare $P_{26|0}$ and $P_{58|2}$ in Figure \ref{bmt_p}. Initially, their probabilities of transitioning from platelet recovery (stage 2) to chronic GVHD (stage 6) and from acute GVHD following platelet recovery (stage 5) to chronic GVHD (stage 8) are quite similar until around the 60th waiting day since entering stage 2 and stage 5. After this time point, the two probabilities start to diverge and the probabilities of transitioning from platelet recovery (stage 2) to chronic GVHD (stage 6) become much higher.

Regarding the FRE estimation results, shown by the dashed lines in Figures \ref{bmt_f} and \ref{bmt_p}, we can observe similar trends to those about the IPCW estimations although the FRE estimations generally yield lower estimated values, which are consistent with the findings from the breast cancer data (see Web Appendix D of Supplementary Material). Notably, the FRE and IPCW estimations for $F_{2|0}$, $P_{25|0}$, and $P_{26|0}$ are quite similar, which can be attributed to the relatively low censoring rate at stage 0, where only 13 patients (approximately 10\% of all patients) were censored. In contrast, there are apparent discrepancies between the IPCW and FRE estimations for $F_{5|2}$ and $P_{58|2}$. These differences can be explained as follows: 1) 54 patients (around 46\% of those entering stage 2) were censored at stage 2, and 2) only 19 transitions (approximately 17\% of those entering stage 2) from stage 2 to stage 5 were observed. The high censoring rate and limited number of observed transitions would contribute to more variability and less reliability in the FRE estimations.

\section{Concluding Remarks}\label{concluding remarks}

In this paper, we propose two nonparametric estimators, the IPCW and FRE estimators, for two important quantities related to stage waiting times conditional on a prior stage visit within a progressive multi-stage model.  The quantities of interest are (1) stage waiting time distribution: the proportion of individuals who stay in stage $j$ within time $t$ of entering stage $j$, and (2) cumulative incidence function: the proportion of individuals who transition from stage $j$ to stage $j'$ within time $t$ of entering stage $j$. The practical utility of studying these quantities has been effectively illustrated by the simulation and case studies presented in Sections \ref{simulation} and \ref{case studies}.

To summarize, there are four advantages of our proposed methodologies: (1) They do not rely on the commonly assumed but often unverified (or violated) Markov model. Thus, our proposed methodologies allow for broader applications to various multi-stage models. This flexibility is achieved by assuming a progressive system with a tree structure. Therefore, quantities conditional on prior stage visits can be explored based on the chain rule of conditional probability and the properties of such a progressive tree structure; (2) Dependent censoring is addressed by using an inverse probability of censoring weighting approach; (3) Our methodologies are based on stage waiting time rather than calendar time, as stage waiting times have more clinical relevance to multi-stage problems; (4) Our estimators are conditional on a past state occupation. This allows us to concentrate on a particular segment of individuals undergoing the multi-stage system without biasing the results. To account for the probability of individuals being censored before transitioning to a future stage of interest, we also explore an alternative nonparametric estimator, the FRE method, using fractional observations. Numerical studies show that while the FRE method produces valid results, the IPCW estimator generally outperforms the FRE estimator across various simulation setups considered in the paper. Therefore, the IPCW estimator is recommended for practical use.

There are still some issues with the proposed nonparametric methods that need to be addressed in future research. For example, the current methods do not account for the potential impact of covariates. To address this, the pseudo-value regression approach introduced by Andersen et al. (2003) could be considered. This method involves deriving pseudo-values, as in the jackknife statistic method, from the marginal stage waiting time distributions (or cumulative incidence functions) and then using these pseudo-values to evaluate the effects of covariates within a generalized estimating equation framework. We plan to investigate the validity of this approach in a future paper. Also, our methodology is currently limited to models with a progressive tree structure. As discussed in Section \ref{methods}, any multi-stage model can be transformed into an expanded model with a progressive tree structure. The progressive tree structure assumption can be relaxed if the conditioning stage in the expanded model is the same as that in the original model. In such cases, the quantities related to stage waiting times in the original model can be derived by summing over the corresponding quantities for the expanded stages. However, when the conditioning stage in the expanded model differs from that in the original model, it is unclear how to generalize the chain rule of probability, which is valid only within the progressive tree structure. Thus, we will explore avenues to extend the product rule approach to broader non-Markov settings. Last but not the least, it should be valuable to extend the proposed methodologies to cases with other forms of incompleteness, such as left truncation, interval censoring, and current status censoring.

\newpage

\begin{center}
   {\bf\Large References}
\end{center}

\begin{description}
\item[] Aalen, O.O. , Borgan, O . and Gjessing, H.K. (2008). {\it Survival and Event History Analysis: A Process Point of View}. New York: Springer-Verlag. 

\item[] Andersen, Per Kragh et al. (1993). {\it Statistical Models Based on Counting Processes}.
Springer Series in Statistics.

\item[] Anyaso-Samuel, Samuel, Dipankar Bandyopadhyay, and Somnath Datta (2023). Pseudo-
value regression of clustered multistate current status data with informative cluster sizes. {\it Statistical Methods in Medical Research}
{\bf 32(8)}, 1494--1510.

\item[] Clahsen, P.C., Thaler, H.T., and Sylvester, R.J. (1996). Improved local control and disease-free survival after perioperative chemotherapy for early-stage breast cancer. A European Organization for Research and Treatment of Cancer Breast Cancer Cooperative Group Study. {\it Journal of Clinical Oncology} {\bf 14(3)}, 745–753.

\item[] Datta, S., and Satten, G.A. (2000). Estimating future stage entry and occupation probabilities in a multistage model based on randomly right-censored data. {\it Statistics \& Probability Letters} {\bf 50(1)}, 89–95.

\item[] Datta, S., and Satten, G.A. (2002). Estimation of integrated transition hazards and stage occupation probabilities for non-Markov systems under dependent censoring. {\it Biometrics} {\bf 58(4)}, 792–802.

\item[] Klein, J.P., and Moeschberger, M.L. (2003). {\it Survival Analysis: Techniques for Censored and Truncated Data} (2nd ed.). New York: Springer Verlag.

\item[] Lan, L., and Datta, S. (2010). Non-parametric estimation of state occupation, entry, and exit times with multistate current status data. {\it Statistical Methods in Medical Research} {\bf 19(2)}, 147–165.

\item[] Lin, D.Y., Sun, W., and Ying, Z. (1999). Nonparametric estimation of the gap time distribution for serial events with censored data. {\it Biometrika}, {\bf 86(1)}, 59–70.

\item[] Mostajabi, F., and Datta, S. (2013). Nonparametric regression of state occupation, entry, exit, and waiting times with multistate right-censored data. {\it Statistics in Medicine}, {\bf 32(17)}, 3006–3019.

\item[] Robins, J.M., and Rotnitzky, A. (1992). Recovery of information and adjustment for dependent censoring using surrogate markers. In: {\it AIDS Epidemiology: Methodological Issues}. Springer, 297–331.

\item[] Satten, G.A., Datta, S., and Robins, J. (2001). Estimating the marginal survival function in the presence of time-dependent covariates. {\it Statistics \& Probability Letters} {\bf 54(4)}, 397–403.

\item[] van der Hage, J. A., van de Velde, C. J., Julien, J. P., Floiras, J. L., Delozier, T., Vandervelden, C., and et al. (2001). Improved survival after one course of perioperative chemotherapy in early breast cancer patients: Long-term results from the European Organization for Research and Treatment of Cancer (EORTC) Trial 10854. {\it European Journal of Cancer} {\bf 37(17)}, 2184–2193.

\item[] Wang, W., and Wells, M.T. (1998). Nonparametric estimation of successive duration times under dependent censoring. {\it Biometrika} {\bf 85(3)}, 561–572.

\item[] Yang, Y., Wu, S., and Datta, S. (2023). Regression analysis of a future state entry time distribution conditional on a past state occupation in a progressive multistate model. {\it Statistical Methods in Medical Research} {\bf 32(12)}, 2285–2298.

\end{description}

\newpage

\section*{Supplementary Material}

\setcounter{equation}{0}
\setcounter{table}{0}
\setcounter{figure}{0}
\renewcommand{\theequation}{A.\arabic{equation}}
\renewcommand{\thetable}{Web Table \arabic{table}}
\renewcommand{\thefigure}{Web Figure \arabic{figure}}
\renewcommand{\tablename}{}  
\renewcommand{\figurename}{} 

\subsection*{Web Appendix A}
\subsubsection*{Proof of Lemma 1}
First define the collections, $\boldsymbol{U_{i}}=(U_{ij}:\ j\ge 1)$, $\boldsymbol{T_{i}}=(T_{ij}:\ j\ge 1)$, $\boldsymbol{\delta_{i}}=(\delta_{ij}:\ j\ge 1)$, and $\boldsymbol{\gamma_{i}}=(\gamma_{ij}:\ j\ge 1)$. Then, define the increasing sigma algebras
\[
   {\mathcal F}_t = \sigma 
   \left( 
   \left\{
\boldsymbol{U_{i}}, \boldsymbol{T_{i}}, \boldsymbol{\delta_{i}}, \boldsymbol{\gamma_{i}}, I\left(T_i\le u, \delta_i=0 \right), 0\le u \le t, i=1, \cdots, n
   \right\}
   \right),\ t\ge 0.
\]
Based on Fubini's theorem, 
\begin{equation}
\begin{aligned}
  \mathbb{E}\left( \bar{N}_{jj'}(t) \right)& = \mathbb{E}\left\{ 
  \displaystyle\sum_{i} \frac{I\left[U_{ij}-T_{ij}\le t, \delta_{ij}=1, \gamma_{ij'}=1\right]}{K_i(U_{ij}-)}
  \right\}\\
  &=n\mathbb{E}\left\{ 
  \frac{I\left[U_{ij}-T_{ij}\le t, \delta_{ij}=1, \gamma_{ij'}=1\right]}{K_i(U_{ij}-)}
  \right\}\\
 &=n\mathbb{E}\left\{ 
  \frac{I\left[U_{ij}-T_{ij}\le t, \delta_{ij}=1, \gamma_{ij'}=1, C_i\ge U_{ij}\right]}{K_i(U_{ij}-)}
  \right\}\\
  \nonumber
\end{aligned}
\end{equation}
Based on Volterra integral and Taylor expansion,
\begin{equation}
I\left[C_i\ge U_{ij} \right]= \displaystyle\prod_{s<U_{ij}}\left[ 1+ {\mathrm d}X_i(s) \right],
\label{X}
\end{equation}
where $X_i(s)=-I\left[C_i\le s\right]=-I\left[C_i\le s, \delta_j=0\right]$. Therefore, 
\begin{equation}
\begin{aligned}
  &I\left[U_{ij}-T_{ij}\le t, \delta_{ij}=1, \gamma_{ij'}=1, C_i\ge U_{ij}\right]\\
  =&I\left[U_{ij}-T_{ij}\le t, \delta_{ij}=1, \gamma_{ij'}=1\right]I\left[ C_i\ge U_{ij}\right]\\
  =&I\left[U_{ij}-T_{ij}\le t, \delta_{ij}=1, \gamma_{ij'}=1\right]\displaystyle\prod_{s<U_{ij}}\left[ 1+ {\mathrm d}X_i(s) \right]\\
  \nonumber
\end{aligned}
\end{equation}

On the set of $\left\{ U_{ij}-T_{ij}\le t, \delta_{ij}=1, \gamma_{ij'}=1, C_i\ge U_{ij} \right\}$, we have
\begin{equation}
K_i(U_{ij}-)=\displaystyle\prod_{s<U_{ij}}\left[ 1+ {\mathrm d}X_i'(s) \right],
\label{X'}
\end{equation}
where $X_i'(s)=-\int_{0}^{U_{ij}} \lambda^c(u)\mathrm{d}u=-\int_{0}^{s} \lambda^c(u) I\left[U_{ij}\ge u\right] \mathrm{d}u$. It should be mentioned that $X_i'(s) - X_i(s)$ is a zero-mean martingale with respect to ${\mathcal F}_t$ and can be written as $M_i^c(s)$.

Based on Duhamel Equation (see Theorem II.6.2 in Andersen et al., 1993) and Equations (\ref{X}) and (\ref{X'}), 
\begin{equation}
\begin{aligned}
  &\frac{I\left[U_{ij}-T_{ij}\le t, \delta_{ij}=1, \gamma_{ij'}=1, C_i\ge U_{ij}\right]}{K_i(U_{ij}-)}\\
  =&I\left[U_{ij}-T_{ij}\le t, \delta_{ij}=1, \gamma_{ij'}=1\right]\frac{I\left[C_i\ge U_{ij}\right]}{K_i(U_{ij}-)}\\
  =&I\left[U_{ij}-T_{ij}\le t, \delta_{ij}=1, \gamma_{ij'}=1\right] \left\{1-\int_{0}^{U_{ij}-}\frac{I\left[C_i\ge s\right]}{K_i(s)}\mathrm{d}M_i^c(s)\right\} \\ 
  =&I\left[U_{ij}-T_{ij}\le t, \delta_{ij}=1, \gamma_{ij'}=1\right]- \\
  &\int_{0}^{U_{ij}-}\frac{I\left[C_i\ge s\right]I\left[U_{ij}-T_{ij}\le t, \delta_{ij}=1, \gamma_{ij'}=1\right]}{K_i(s)}\mathrm{d}M_i^c(s) \\
  =&I\left[U_{ij}-T_{ij}\le t, \delta_{ij}=1, \gamma_{ij'}=1\right]-\\
  &\int_{0}^{t}\frac{I\left[U_{ij}-T_{ij}\le t, \delta_{ij}=1, \gamma_{ij'}=1, U_{ij}\ge s\right]}{K_i(s)}\mathrm{d}M_i^c(s), \\
\end{aligned}
\label{duhamel}
\end{equation}
where the second term on the right-hand side of the final equation in (\ref{duhamel}) is also a zero-mean martingale with respect to ${\mathcal F}_t$, as $I\left[U_{ij}-T_{ij}\le t, \delta_{ij}=1, \gamma_{ij'}=1, U_{ij}\ge s\right]/K_i(s)$ is predictable with respect to ${\mathcal F}_s$. Thus, 
\begin{equation}
\begin{aligned}
  &\mathbb{E}\left\{ 
  I\left[U_{ij}-T_{ij}\le t, \delta_{ij}=1, \gamma_{ij'}=1, C_i\ge U_{ij}\right]
  \right\}\\
  =&\mathbb{E}\left\{ 
  \frac{I\left[U_{ij}-T_{ij}\le t, \delta_{ij}=1, \gamma_{ij'}=1, C_i\ge U_{ij}\right]}{K_i(U_{ij}-)}
  \right\}.
  \nonumber
\end{aligned}
\end{equation}
It should be mentioned that $I\left[U_{ij}^*-T_{ij}^*\le t,\ X^*_{ij}=1,\ X^*_{ij'}=1\right]$ in uncensored cases (see Formula (3) for $N^*_{jj'}$) is equivalent to  $I\left[U_{ij}-T_{ij}\le t, \delta_{ij}=1, \gamma_{ij'}=1, C_i\ge U_{ij}\right]$ since $C_i$ can be treated as infinity in the uncensored cases. Therefore, we prove that 
\[
\mathbb{E}[N^*_{jj'}(t)]=\mathbb{E}[\bar{N}_{jj'}(t)].
\]
Similarly, $\mathbb{E}[N^*_{j}(t)]=\mathbb{E}[\bar{N}_{j}(t)]$ and $\mathbb{E}[Y^*_{j}(t)]=\mathbb{E}[\bar{Y}_{j}(t)]$ can be proved.

\subsection*{Web Appendix B}
\subsubsection*{Proof of Theorem 1}
To prove the consistency of $\hat{S}_j(t)=\displaystyle\prod_{s<t}\left[ 1- \hat{\Lambda}_j({\mathrm d}s) \right]$ for estimating $S_j(t)$, we first demonstrate the consistency of $\hat{\Lambda}_j(t)$ for estimating $\Lambda_j(t)$. 

Assume $t$ is such that $y_j(t):=\Pr(U_{ij}^*-T_{ij}^*\le t, X_{ij}=1)>0$ and $\displaystyle \sup_i\left\{\mathbb{E}[K_i^{-2}(t)]\right\}<\infty$. For simplicity, suppose all variables follow continuous distributions. Using the uniform Cesàro consistency of  $\hat{K}_i(t)$ within compact sets under Aalen’s linear model, follows directly that
\begin{equation}
    \frac{\hat{Y}_j(t)}{n}=\frac{\bar{Y}_j(t)}{n}+o_p(1)\xrightarrow{P} y_j(t),
    \label{conv: Y_hat}
\end{equation}
where, by the law of large numbers for i.i.d. variables and the equivalence $\mathbb{E}[Y^*_{j}(t)]=\mathbb{E}[\bar{Y}_{j}(t)]$, we have
\begin{equation}
    \frac{Y_j^*(t)}{n}\xrightarrow{P} y_j(t).
    \label{conv: Y_star}
\end{equation}
Similarly, assuming $S_j(t):=\Pr(U_{ij}^*-T_{ij}^*\ge t, X_{ij}=1)$ and $S_{jj'}(t):=\Pr(U_{ij}^*-T_{ij}^*\ge t, X_{ij}=1, X_{ij'}=1)$, we have
\begin{equation}
    \frac{\hat{N}_j(t)}{n}\xrightarrow{P} 1-S_j(t),\ \frac{N_j^*(t)}{n}\xrightarrow{P} 1-S_j(t),
    \label{conv: N_j}
\end{equation}
\begin{equation}
    \frac{\hat{N}_{jj'}(t)}{n}\xrightarrow{P} 1-S_{jj'}(t),\ \frac{N_{jj'}^*(t)}{n}\xrightarrow{P} 1-S_{jj'}(t).
    \label{conv: N_jj'}
\end{equation}

The Nelson–Aalen estimator for $\Lambda_j(t)$ is defined as
\[
\Tilde{\Lambda}_j(t)=\int_{0}^{t}J_j^*(s)\mathrm{d}\Lambda_j(s)=\int_{0}^{t}\frac{J_j^*}{Y_j^*(s)}\mathrm{d}\left[N_j^*(s)-M_j^*(s)\right],
\]
where $J_j^*(t)=I\left[Y_j^*(t)>0\right]$. It should be noted that $\Tilde{\Lambda}_j(t)=\Lambda_j(t)+o_p(1)$ under the condition that $y_j(t)>0$. Using (\ref{conv: Y_hat})-(\ref{conv: N_j}), we can derive the following expression:
\begin{equation}
\begin{aligned}
  \left|\hat{\Lambda}_j(t)-\Tilde{\Lambda}_j(t)\right|&=
  \left|\int_{0}^{t}J_j^*(s)\left\{ \frac{\mathrm{d}\hat{N}_j(s)}{\hat{Y}_j(s)} - \frac{\mathrm{d}\left[N_j^*(s)-M_j^*(s)\right]}{Y_j^*(s)} \right\}\right|\\
  & = \left|\int_{0}^{t}\frac{J_j^*(s)}{Y_j^*(s)}\mathrm{d}M_j^*(s)+ \int_{0}^{t}J_j^*(s)\left\{ \frac{\mathrm{d}\hat{N}_j(s)}{\hat{Y}_j(s)} - \frac{\mathrm{d}N_j^*(s)}{Y_j^*(s)} \right\}\right|\\
  & \le  \left|\int_{0}^{t}\frac{J_j^*(s)}{Y_j^*(s)}\mathrm{d}M_j^*(s)\right|
  +\int_{0}^{t}\left|\frac{J_j^*(s)}{\hat{Y}_j(s)/n}-\frac{1}{y_j(s)}\right|\frac{\mathrm{d}\hat{N}_j(s)}{n}\\
  &\ \ \  + \int_{0}^{t}\left|\frac{1}{y_j(s)}-\frac{J_j^*(s)}{Y^*_j(s)/n}\right|\frac{\mathrm{d}N^*_j(s)}{n}+ \int_{0}^{t} \frac{1}{y_j(s)} \left| \frac{\mathrm{d}\hat{N}_j(s)}{n}-\frac{\mathrm{d}N^*_j(s)}{n} \right|\\
  & \xrightarrow{P} 0,\ \text{as}\ n\xrightarrow{} \infty.
  \label{conv: Lambda}
\end{aligned}
\end{equation}
Therefore we have $\hat{\Lambda}_j(t)$ is a consistent estimator of $\Lambda_j(t)$. Following continuity results from the Duhamel equation, we conclude that $\hat{S}_j(t)$ is a consistent estimator of $S_j(t)$. 

Next, we show the proof of the consistency of $\boldsymbol{\hat{P}}(t)$ for estimating $\boldsymbol{P}(t)$. Firstly, we define the transition probability matrix as 
\begin{equation}
    \boldsymbol{P}(t)=\displaystyle\prod_{s<t}\left[ I- {\mathrm d}\boldsymbol{A}(s) \right],
\end{equation}
where $\boldsymbol{A}(s)$ represents the integrated transition hazard matrix. Here, $P_{jj'}(t)$ is the ($j$, $j'$)th element of $\boldsymbol{P}(t)$, representing the transition probability (i.e., cumulative incidence function) from stage $j$ to stage $j'$ at time $t$. Similarly, to prove the consistency of $\boldsymbol{\hat{P}}(t)$ for $\boldsymbol{P}(t)$, we start with the proof of the consistency of $\boldsymbol{\hat{A}}(t)$ for $\boldsymbol{A}(t)$.

Assume that $j$ is a non-terminal stage with $j \neq j'$. The Nelson-Aalen estimator for the $(j, j')$th element of $\boldsymbol{A}(t)$, denoted $A_{jj'}$, is defined as
\[
\Tilde{A}_{jj'}(t)=\int_{0}^{t}J_j^*(s)\mathrm{d}A_{jj'}(s)=\int_{0}^{t}\frac{J_j^*}{Y_j^*(s)}\mathrm{d}\left[N_{jj'}^*(s)-M_{jj'}^*(s)\right],
\]
which satisfies $\Tilde{A}_{jj'}(t)=A_{jj'}(t)+o_p(1)$ under the condition that $y_j(t) > 0$. Following the approach used in deriving consistency in (\ref{conv: Lambda}) and using (\ref{conv: Y_hat}), (\ref{conv: Y_star}) and (\ref{conv: N_jj'}), we have
\begin{equation}
\begin{aligned}
  \left|\hat{A}_{jj'}(t)-\Tilde{A}_{jj'}(t)\right|&=
  \left|\int_{0}^{t}J_j^*(s)\left\{ \frac{\mathrm{d}\hat{N}_{jj'}(s)}{\hat{Y}_j(s)} - \frac{\mathrm{d}\left[N_{jj'}^*(s)-M_{jj'}^*(s)\right]}{Y_j^*(s)} \right\}\right|\\
  & = \left|\int_{0}^{t}\frac{J_j^*(s)}{Y_j^*(s)}\mathrm{d}M_{jj'}^*(s)+ \int_{0}^{t}J_j^*(s)\left\{ \frac{\mathrm{d}\hat{N}_{jj'}(s)}{\hat{Y}_j(s)} - \frac{\mathrm{d}N_{jj'}^*(s)}{Y_j^*(s)} \right\}\right|\\
  & \le  \left|\int_{0}^{t}\frac{J_j^*(s)}{Y_j^*(s)}\mathrm{d}M_{jj'}^*(s)\right|
  +\int_{0}^{t}\left|\frac{J_j^*(s)}{\hat{Y}_j(s)/n}-\frac{1}{y_j(s)}\right|\frac{\mathrm{d}\hat{N}_{jj'}(s)}{n}\\
  &\ \ \  + \int_{0}^{t}\left|\frac{1}{y_j(s)}-\frac{J_j^*(s)}{Y^*_j(s)/n}\right|\frac{\mathrm{d}N^*_{jj'}(s)}{n}+ \int_{0}^{t} \frac{1}{y_j(s)} \left| \frac{\mathrm{d}\hat{N}_{jj'}(s)}{n}-\frac{\mathrm{d}N^*_{jj'}(s)}{n} \right|\\
  & \xrightarrow{P} 0,\ \text{as}\ n\xrightarrow{} \infty.
  \label{conv: Lambda}
\end{aligned}
\end{equation}
Therefore, we conclude that $\hat{A}_{jj'}(t)$ is a consistent estimator of $A_{jj'}(t)$. From the simple continuity result from the Duhamel equation, we establish that $\hat{P}_{jj'}(t)$ is a consistent estimator of $P_{jj'}(t)$. 

\subsection*{Web Appendix C}
\subsubsection*{Additional results of simulation studies}

\begin{figure}[H]
    \centering
    \begin{subfigure}[t]{0.45\textwidth}
        \centering
        \includegraphics[scale=.55]{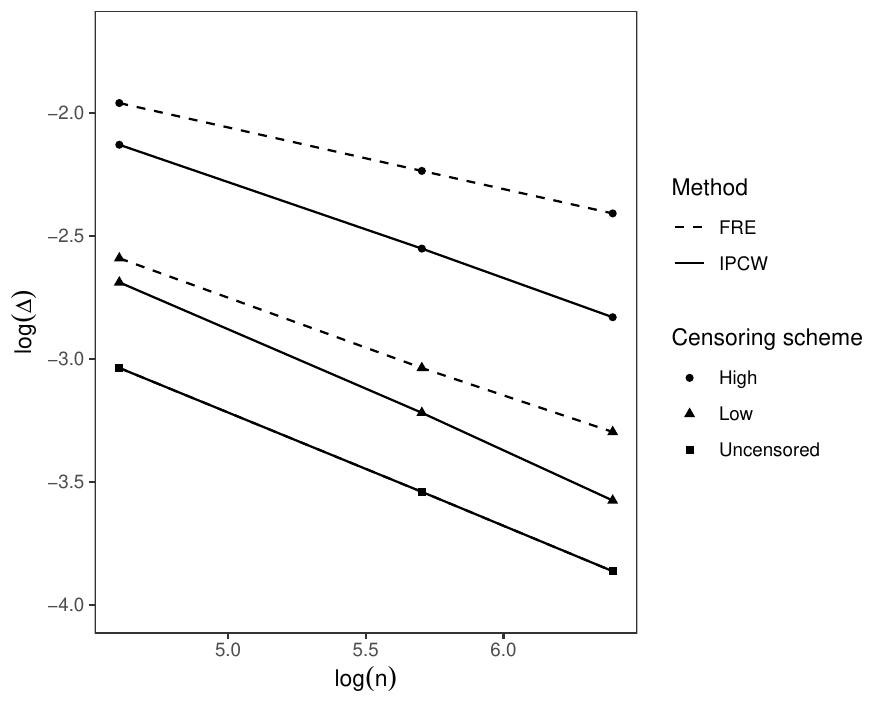}
        \caption{WB waiting and censoring times; independent censoring} 
    \end{subfigure}
    \hfill
    \begin{subfigure}[t]{0.45\textwidth}
        \centering
        \includegraphics[scale=.55]{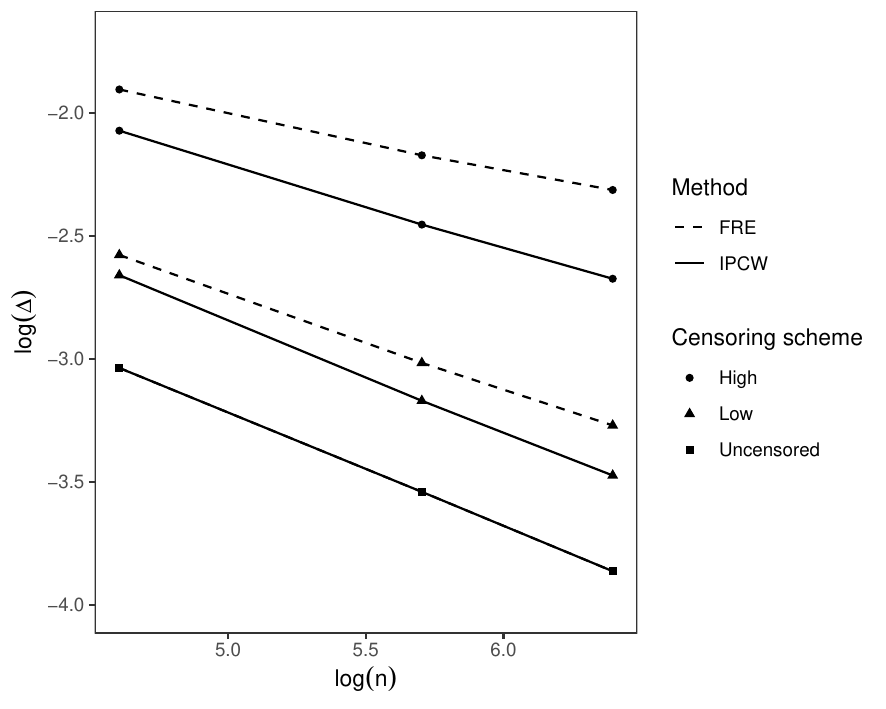} 
        \caption{WB waiting and censoring times; stage-dependent censoring} 
    \end{subfigure}

    \vspace{1cm}
    \begin{subfigure}[t]{0.45\textwidth}
        \centering
        \includegraphics[scale=.55]{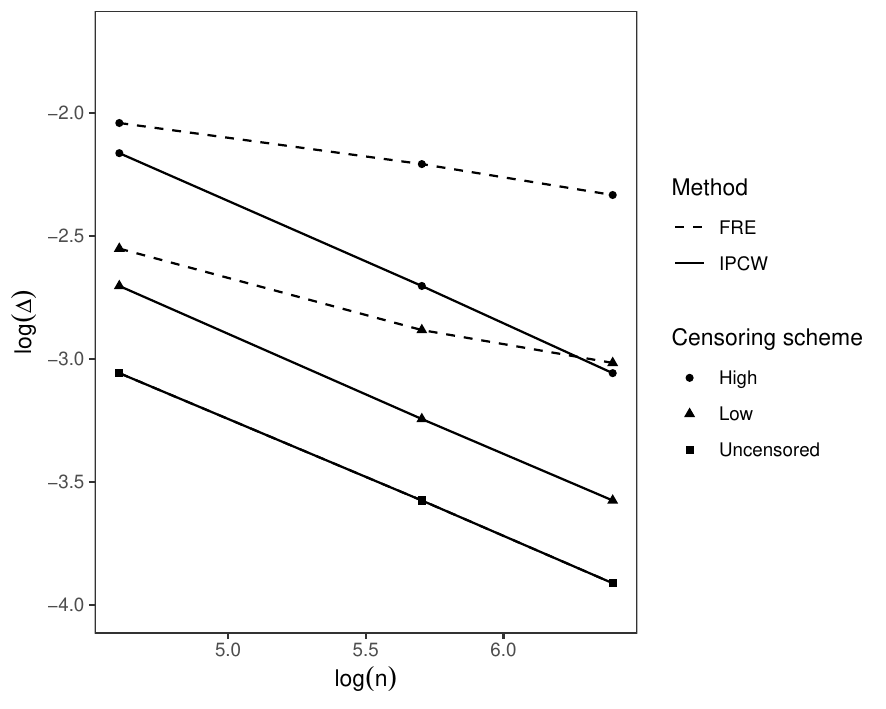}
        \caption{LN waiting and censoring times; independent censoring} 
    \end{subfigure}
    \hfill
    \begin{subfigure}[t]{0.45\textwidth}
        \centering
        \includegraphics[scale=.55]{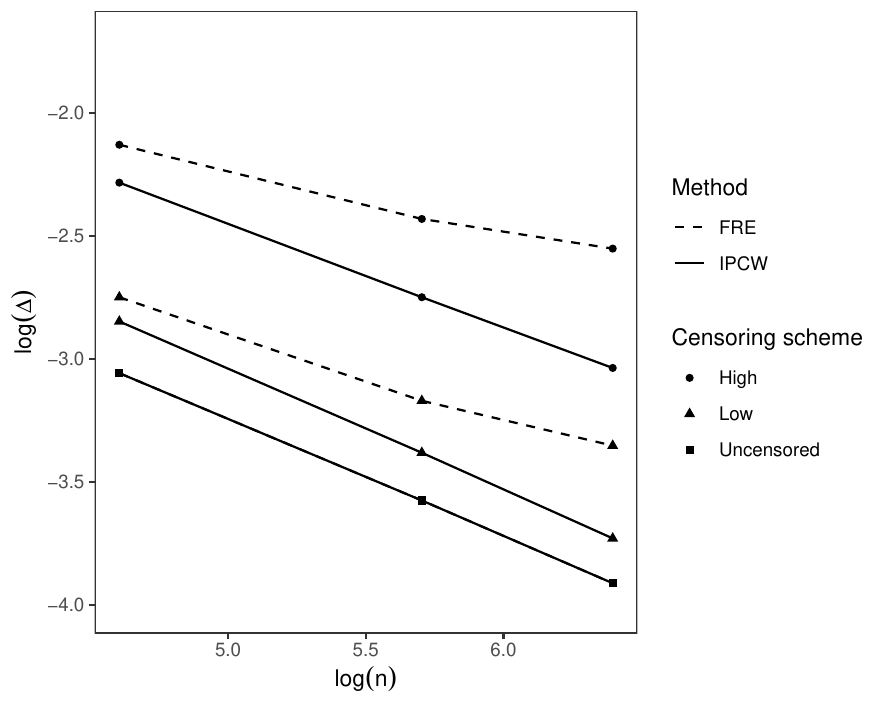}
        \caption{LN waiting and censoring times; stage-dependent censoring} 
    \end{subfigure}
    \caption{Relationships between the logarithms of the $L_1$ norm of the estimation errors and the logarithms of the sample size for the IPCW and FRE estimators of $P_{35|1}$ under the Markov model.}
    \label{fig:L1:P:Markov}
\end{figure}

\begin{figure}[H]
    \centering
    \begin{subfigure}[t]{0.45\textwidth}
        \centering
        \includegraphics[scale=.55]{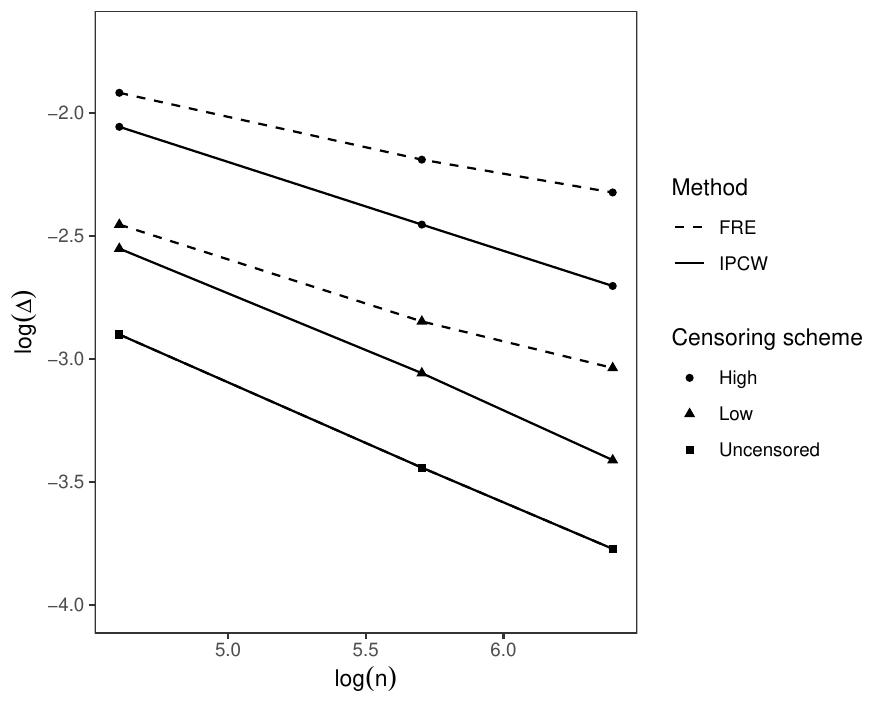}
        \caption{WB waiting and censoring times; independent censoring} 
    \end{subfigure}
    \hfill
    \begin{subfigure}[t]{0.45\textwidth}
        \centering
        \includegraphics[scale=.55]{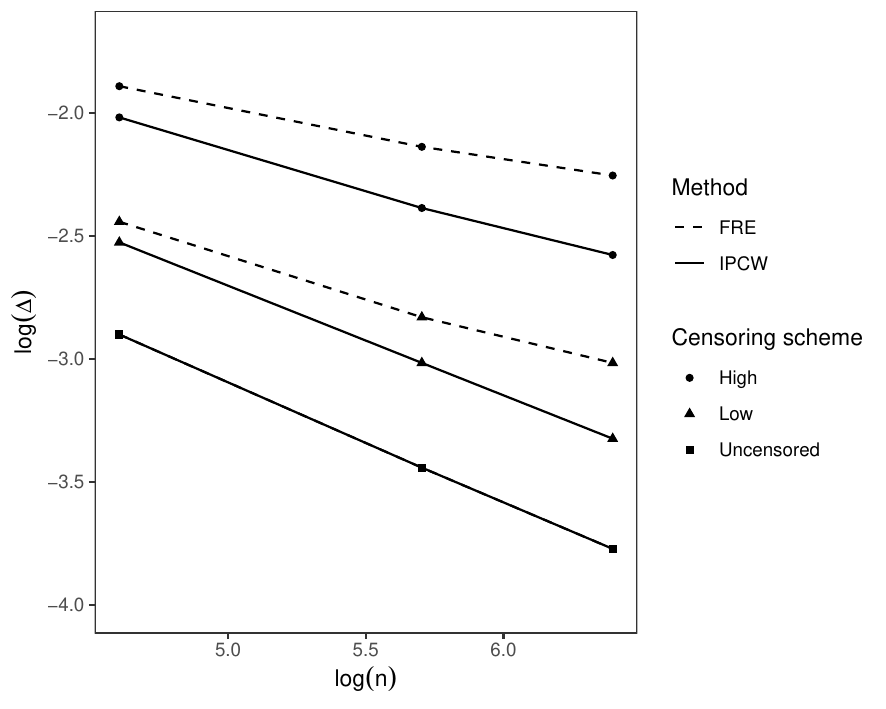} 
        \caption{WB waiting and censoring times; stage-dependent censoring} 
    \end{subfigure}

    \vspace{1cm}
    \begin{subfigure}[t]{0.45\textwidth}
        \centering
        \includegraphics[scale=.55]{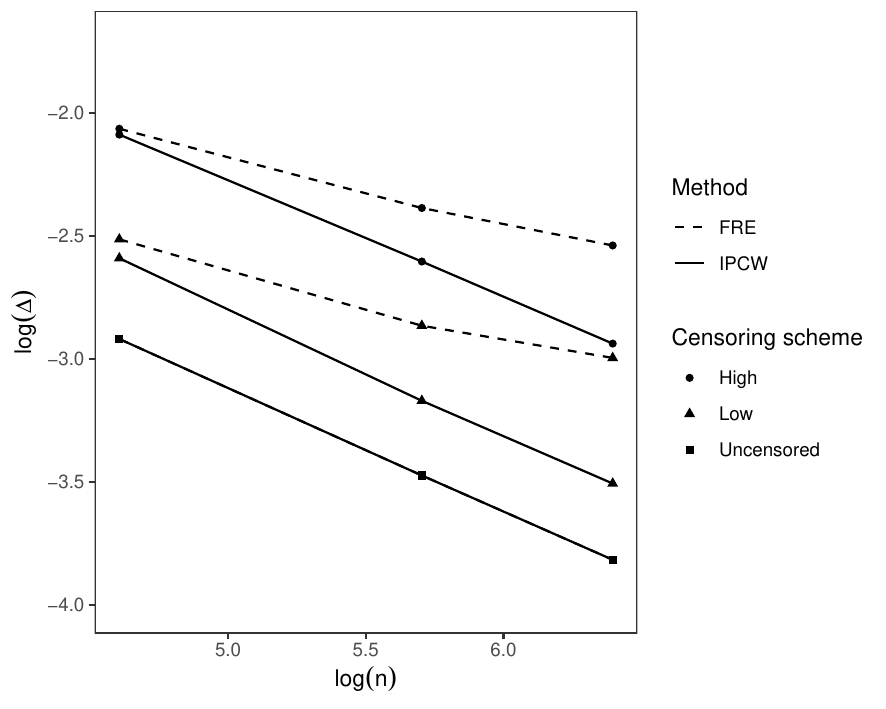}
        \caption{LN waiting and censoring times; independent censoring} 
    \end{subfigure}
    \hfill
    \begin{subfigure}[t]{0.45\textwidth}
        \centering
        \includegraphics[scale=.55]{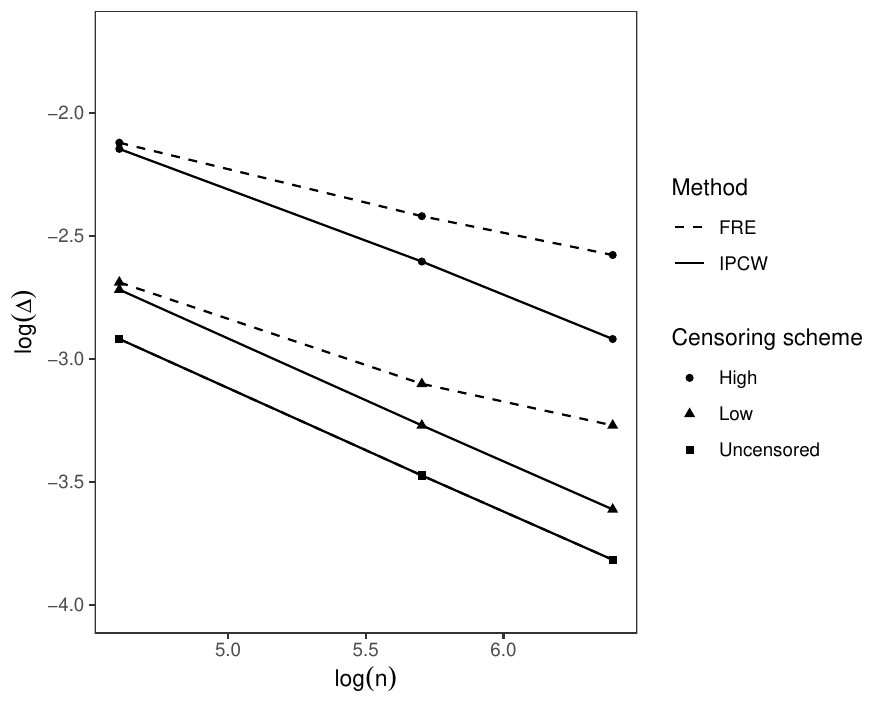}
        \caption{LN waiting and censoring times; stage-dependent censoring} 
    \end{subfigure}
    \caption{Relationships between the logarithms of the $L_1$ norm of the estimation errors and the logarithms of the sample size for the IPCW and FRE estimators of $F_{3|1}$ under the semi-Markov model.}
    \label{fig:L1:F:semi}
\end{figure}

\begin{figure}[H]
    \centering
    \begin{subfigure}[t]{0.45\textwidth}
        \centering
        \includegraphics[scale=.55]{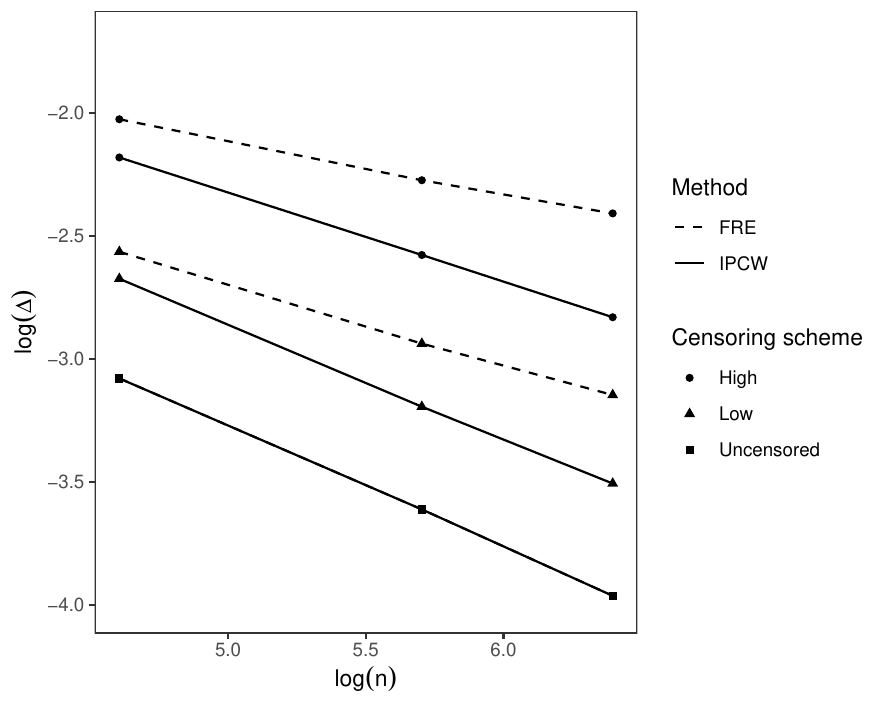}
        \caption{WB waiting and censoring times; independent censoring} 
    \end{subfigure}
    \hfill
    \begin{subfigure}[t]{0.45\textwidth}
        \centering
        \includegraphics[scale=.55]{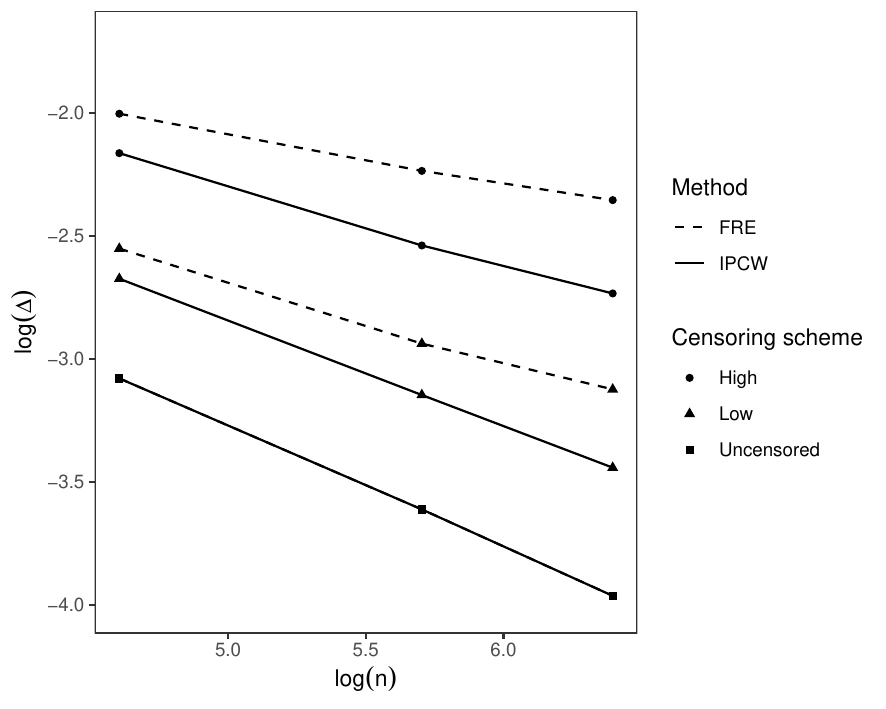} 
        \caption{WB waiting and censoring times; stage-dependent censoring} 
    \end{subfigure}

    \vspace{1cm}
    \begin{subfigure}[t]{0.45\textwidth}
        \centering
        \includegraphics[scale=.55]{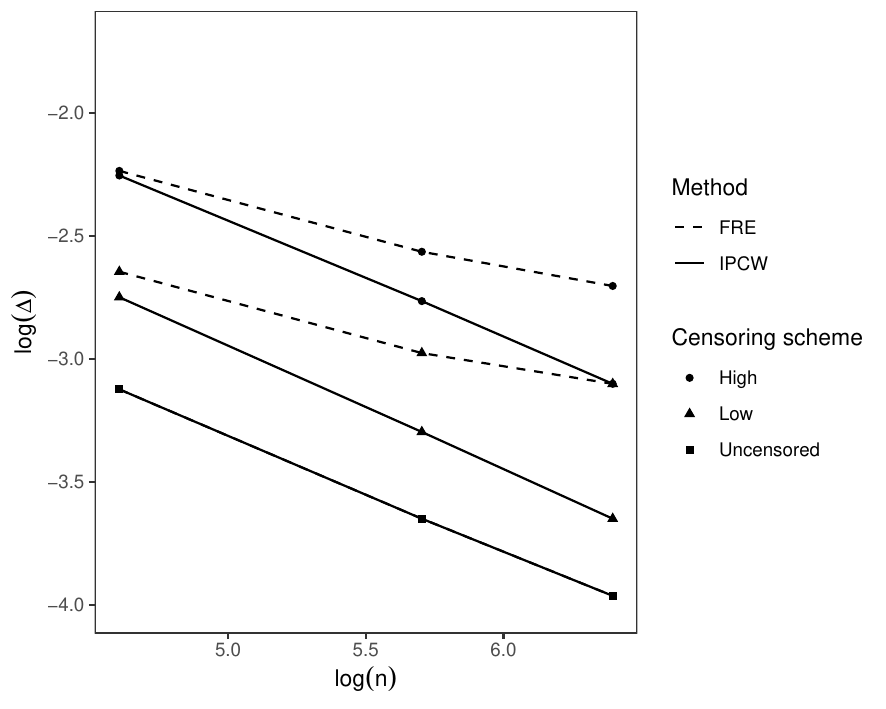}
        \caption{LN waiting and censoring times; independent censoring} 
    \end{subfigure}
    \hfill
    \begin{subfigure}[t]{0.45\textwidth}
        \centering
        \includegraphics[scale=.55]{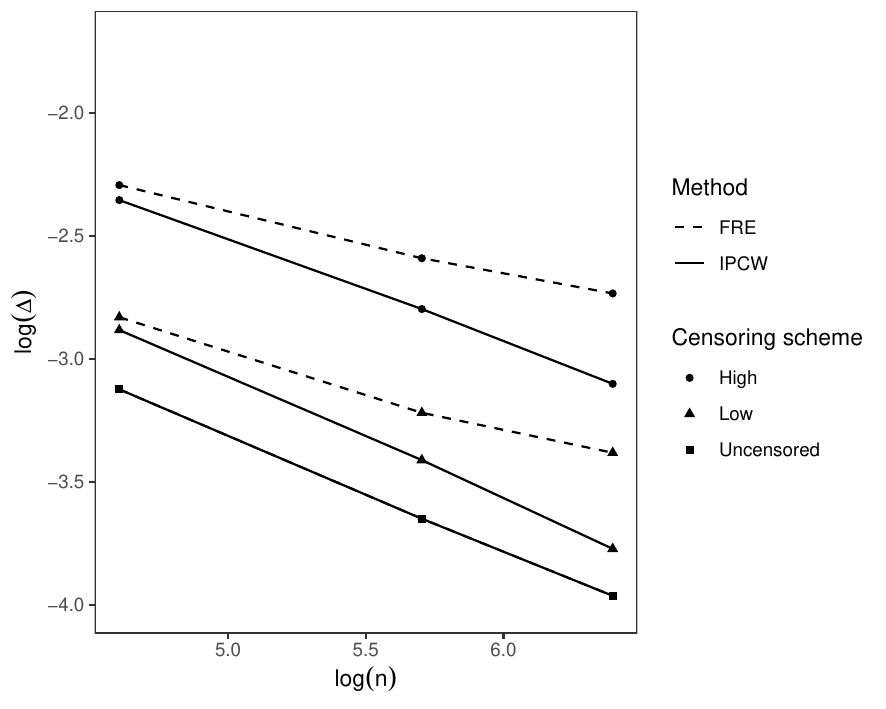}
        \caption{LN waiting and censoring times; stage-dependent censoring} 
    \end{subfigure}
    \caption{Relationships between the logarithms of the $L_1$ norm of the estimation errors and the logarithms of the sample size for the IPCW and FRE estimators of $P_{35|1}$ under the semi-Markov model.}
    \label{fig:L1:P:semi}
\end{figure}

To better demonstrate the efficacy of our proposed estimators, we examine the relationships between the logarithms of the $L_1$ norm of the estimation errors and the logarithms of the sample size $n$ for the IPCW and FRE estimators of $F_{3|1}(t)$ and $P_{35|1}(t)$ under different simulation settings. In the main paper, due to page limitations, we present only Figure 3, which shows approximately linear relationships between the logarithms of the $L_1$ norm of the estimation errors and the logarithms of the sample size $n$ for each estimator of $F_{3|1}(t)$ under different simulation settings for the Markov model. Web Figures~\ref{fig:L1:P:Markov}--\ref{fig:L1:P:semi} present the corresponding results for each estimator of $F_{3|1}(t)$ and $P_{35|1}(t)$ under other simulation settings considered in the main paper.

\begin{figure}[H]
    \centering
    \begin{subfigure}[t]{0.45\textwidth}
        \centering
        \includegraphics[scale=.55]{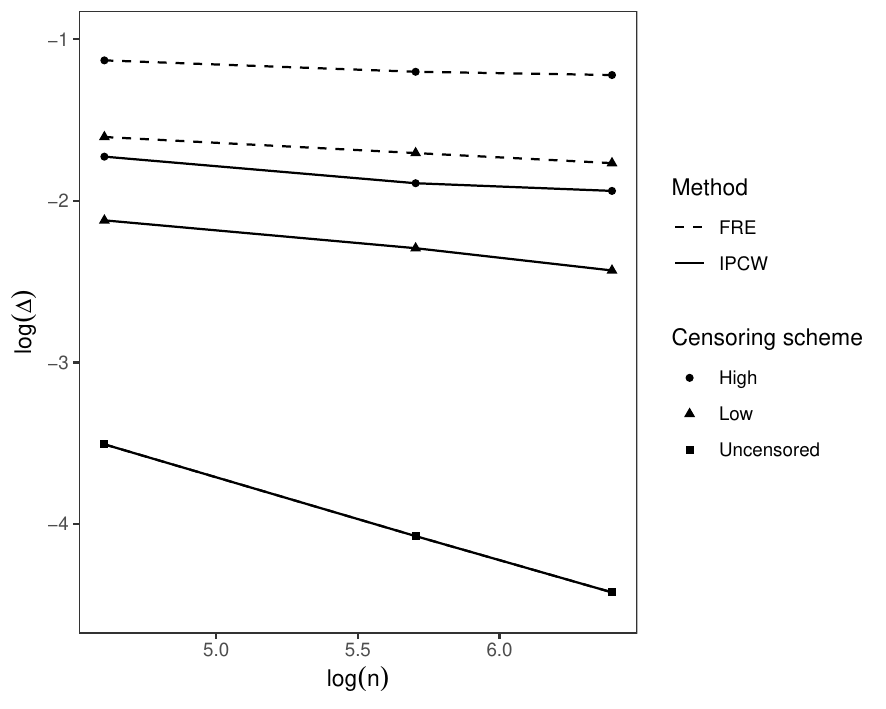}
        \caption{WB waiting and censoring times; independent censoring} 
    \end{subfigure}
    \hfill
    \begin{subfigure}[t]{0.45\textwidth}
        \centering
        \includegraphics[scale=.55]{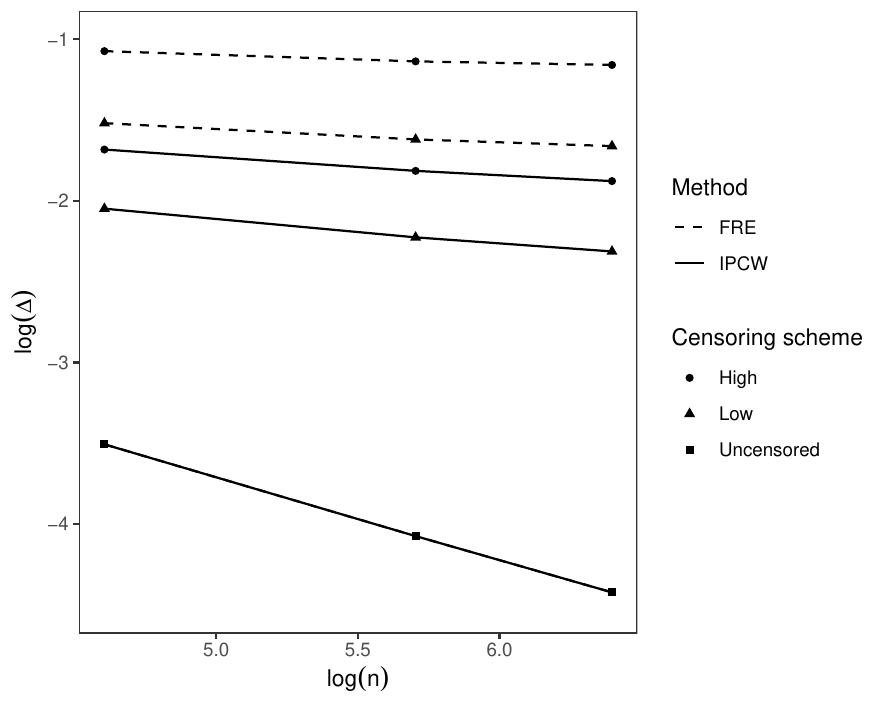} 
        \caption{WB waiting and censoring times; stage-dependent censoring} 
    \end{subfigure}

    \vspace{1cm}
    \begin{subfigure}[t]{0.45\textwidth}
        \centering
        \includegraphics[scale=.55]{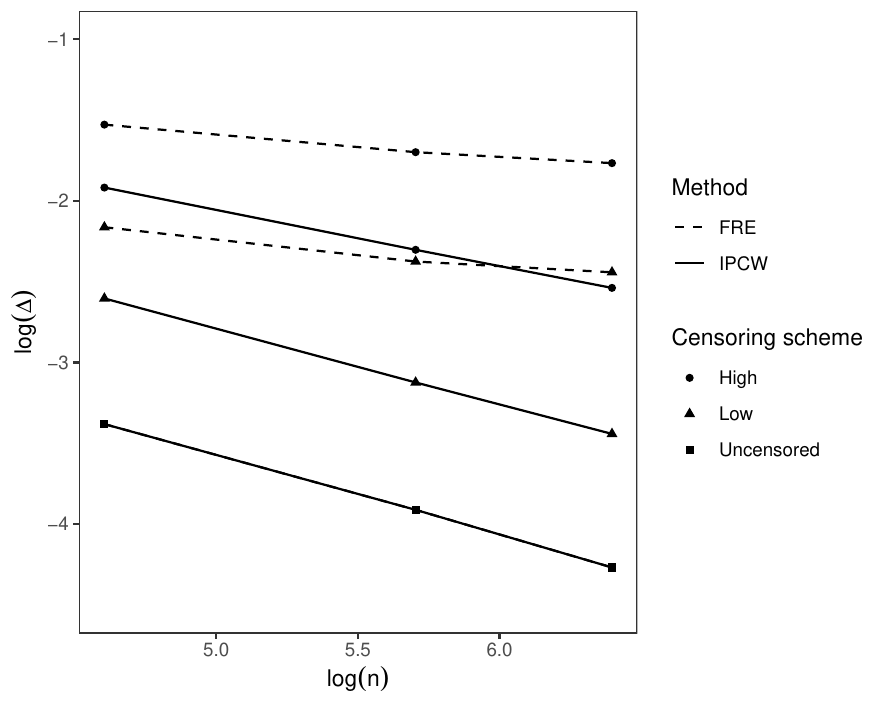}
        \caption{LN waiting and censoring times; independent censoring} 
    \end{subfigure}
    \hfill
    \begin{subfigure}[t]{0.45\textwidth}
        \centering
        \includegraphics[scale=.55]{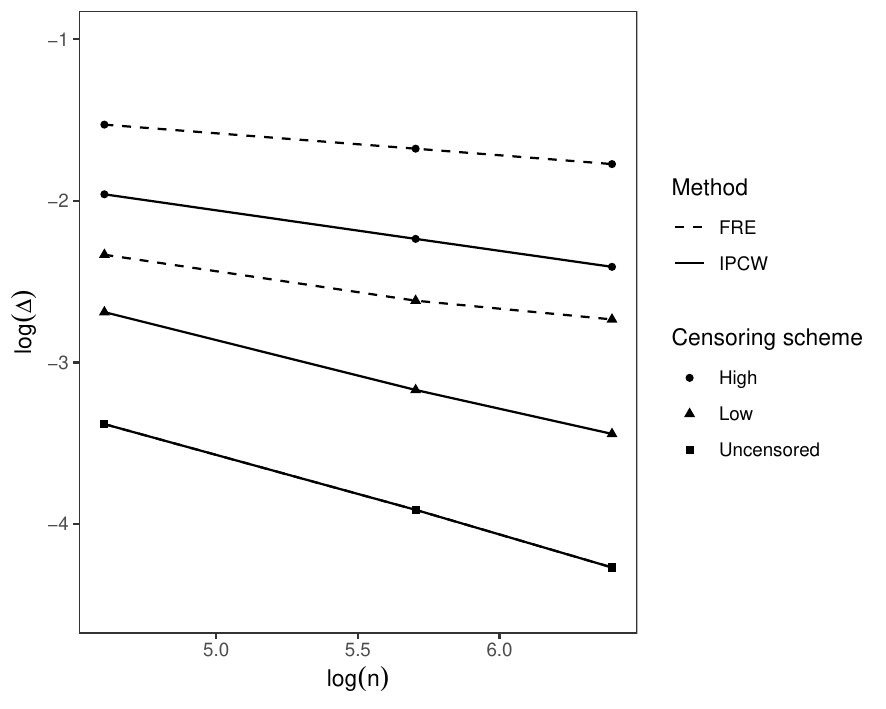}
        \caption{LN waiting and censoring times; stage-dependent censoring} 
    \end{subfigure}
    \caption{Relationships between the logarithms of the $L_1$ norm of the estimation errors and the logarithms of the sample size for the IPCW and FRE estimators of $F_{3|1}$ under the additional simulation setups of the semi-Markov model with $\alpha = \beta = 1$ and $\tau = 1$.}
    \label{fig:L1:F:semi:extra}
\end{figure}

\begin{figure}[H]
    \centering
    \begin{subfigure}[t]{0.45\textwidth}
        \centering
        \includegraphics[scale=.55]{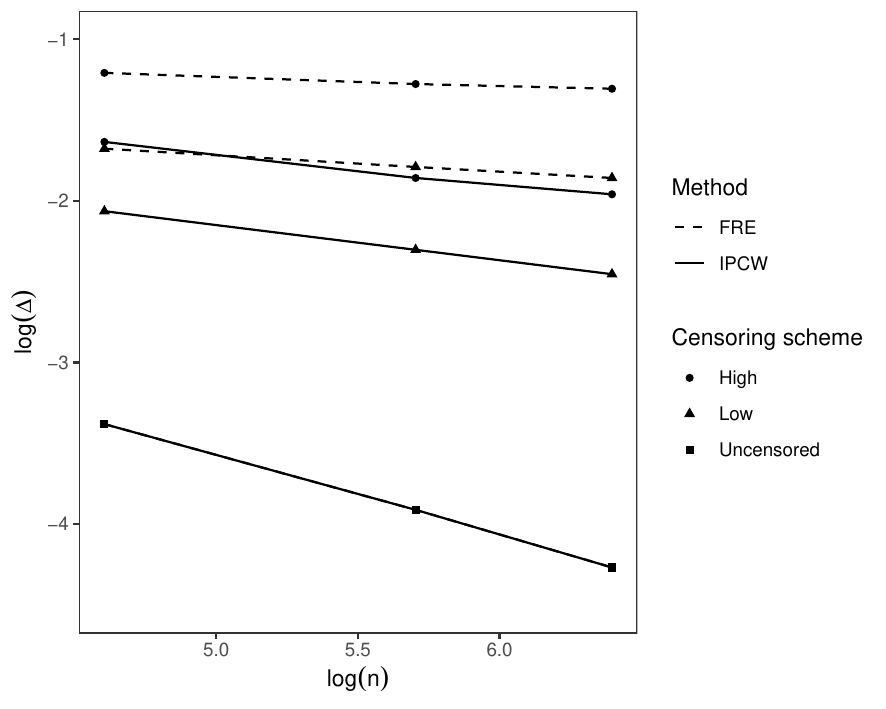}
        \caption{WB waiting and censoring times; independent censoring} 
    \end{subfigure}
    \hfill
    \begin{subfigure}[t]{0.45\textwidth}
        \centering
        \includegraphics[scale=.55]{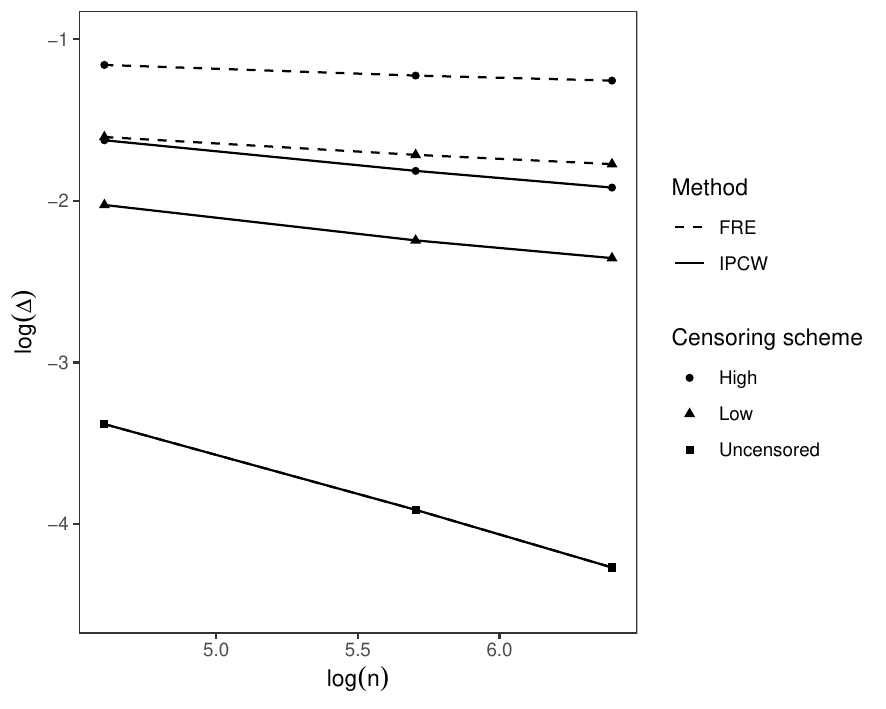} 
        \caption{WB waiting and censoring times; stage-dependent censoring} 
    \end{subfigure}

    \vspace{1cm}
    \begin{subfigure}[t]{0.45\textwidth}
        \centering
        \includegraphics[scale=.55]{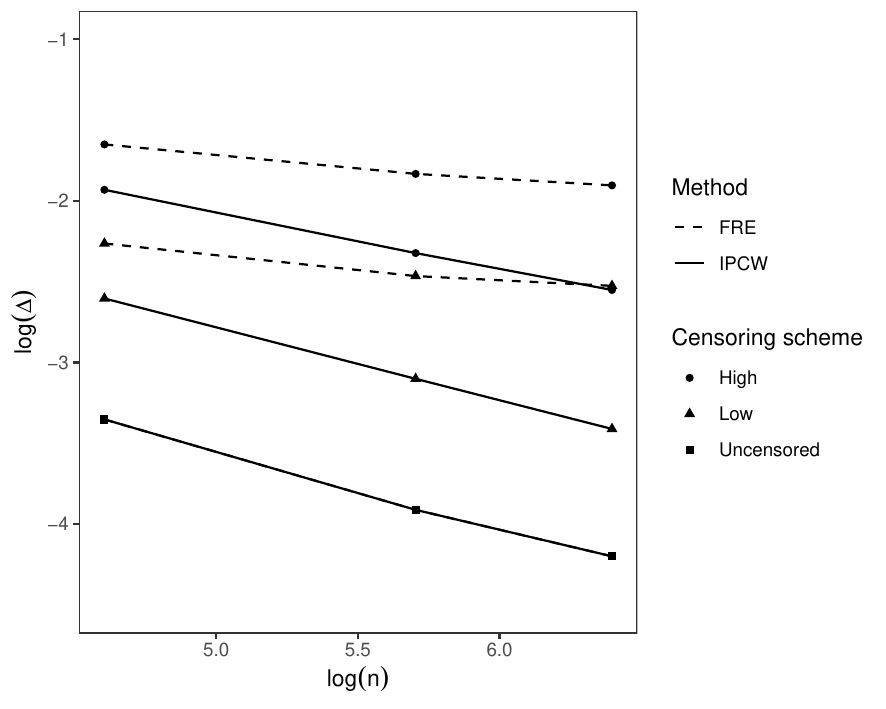}
        \caption{LN waiting and censoring times; independent censoring} 
    \end{subfigure}
    \hfill
    \begin{subfigure}[t]{0.45\textwidth}
        \centering
        \includegraphics[scale=.55]{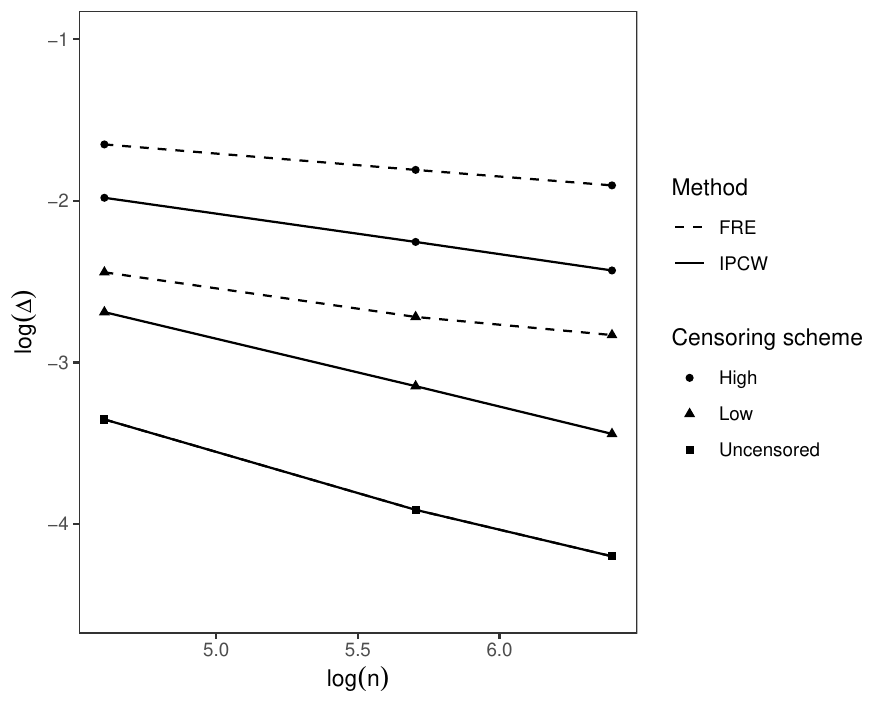}
        \caption{LN waiting and censoring times; stage-dependent censoring} 
    \end{subfigure}
    \caption{Relationships between the logarithms of the $L_1$ norm of the estimation errors and the logarithms of the sample size for the IPCW and FRE estimators of $P_{35|1}$ under the additional simulation setups of the semi-Markov model with $\alpha = \beta = 1$ and $\tau = 1$.}
    \label{fig:L1:P:semi:extra}
\end{figure}

We also perform an additional simulation under the semi-Markov model with $\alpha = \beta = 1$ and $\tau = 1$. All other simulation settings, including the waiting time distribution, the censoring time distribution, the censoring rate, and the censoring scheme, remain consistent with those used in the main paper. Web Figures~\ref{fig:L1:F:semi:extra} and \ref{fig:L1:P:semi:extra} show that the logarithms of the $L_1$ norm of the estimation errors decrease approximately linearly as the logarithms of the sample size increase for both the IPCW and FRE estimators of $F_{3|1}(t)$ and $P_{35|1}(t)$ under this additional setting. Given the large number of possible choices for $\alpha$, $\beta$, and $\tau$, we present results only for the case where $\alpha = \beta = 1$ and $\tau = 1$. Results for other combinations of the parameters are available upon request.

\begin{figure}[H]
\centering
\includegraphics[scale=.70]{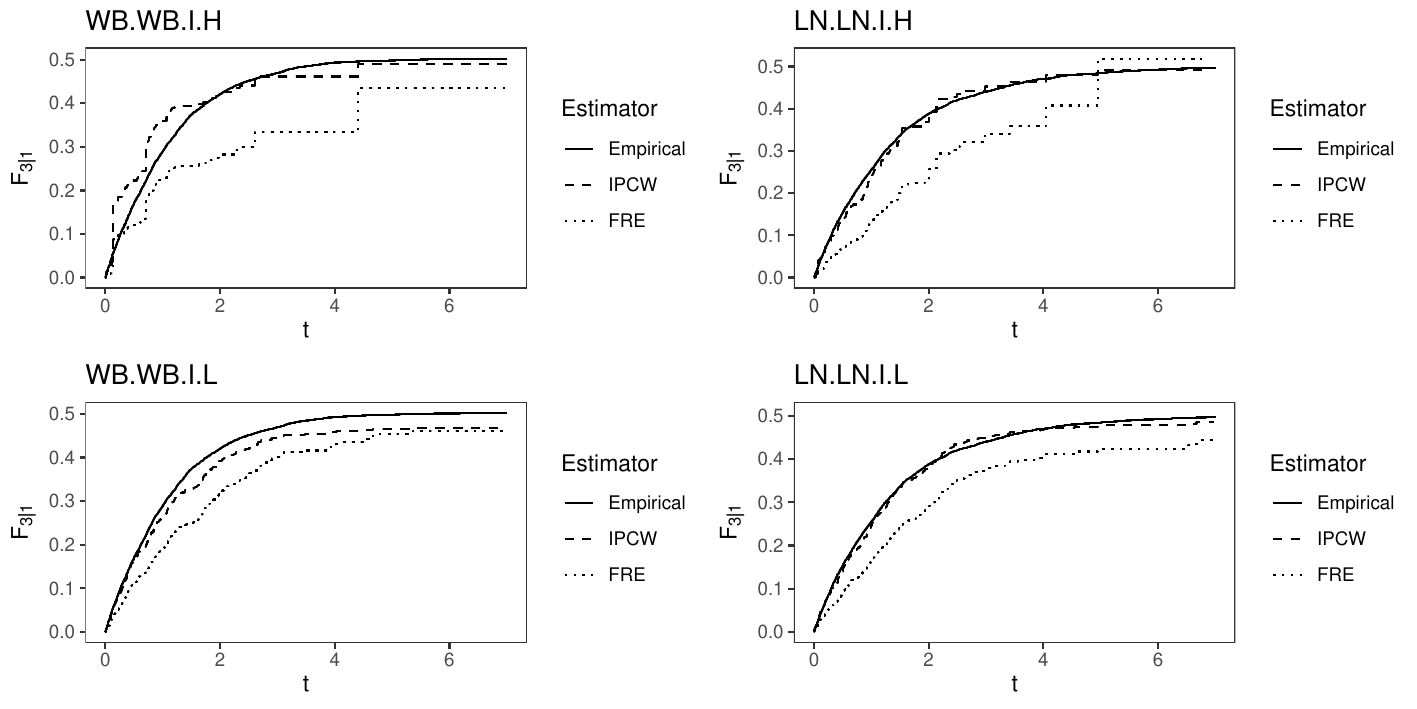}
\caption{Comparison of IPCW and FRE estimators (based on a sample size of 2000 under a specific simulation design) with the empirical true distribution (based on a sample size of 10000 under the simulation design with no censoring) for $F_{3|1}(t)$ in the six-stage Markov models under independent censoring.}
\label{Markov_I_F}
\end{figure}

\begin{figure}[H]
\centering
\includegraphics[scale=.70]{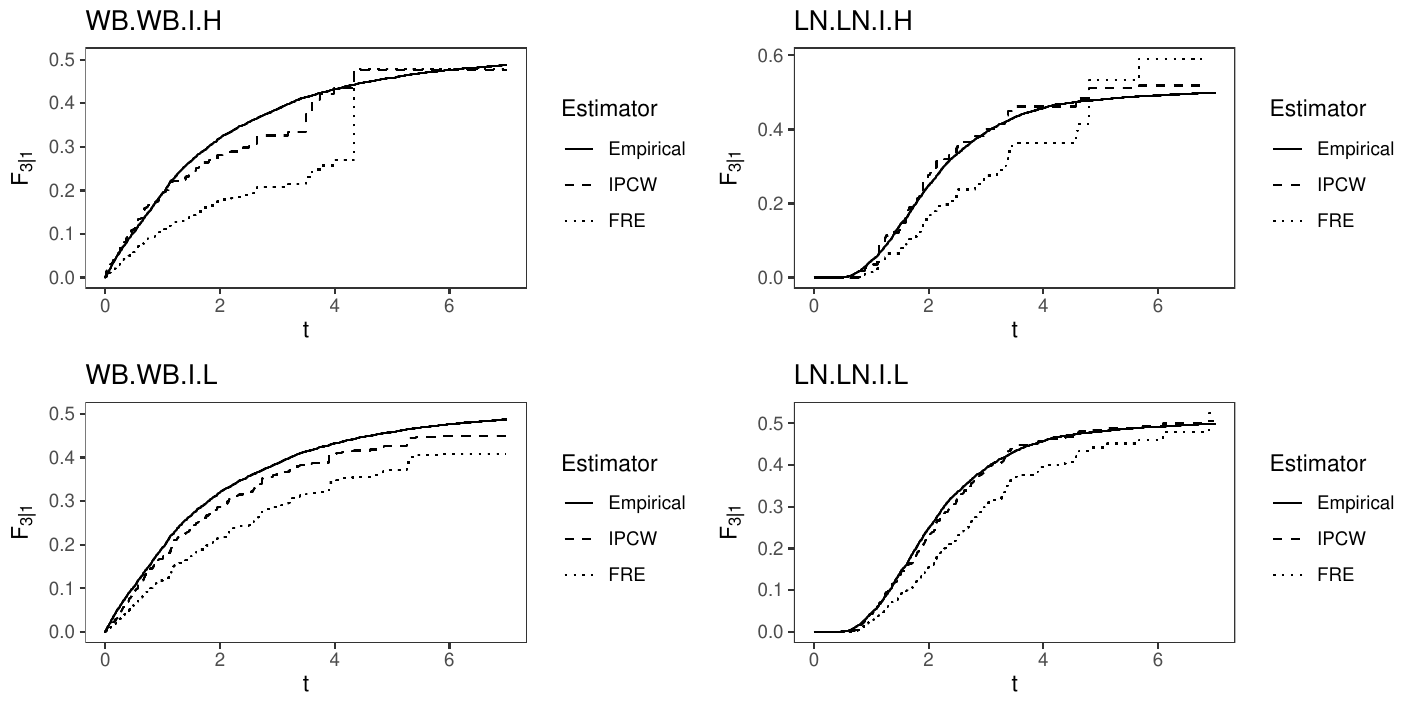}
\caption{Comparison of IPCW and FRE estimators (based on a sample size of 2000 under a specific simulation design) with the empirical true distribution (based on a sample size of 10000 under the simulation design with no censoring) for $F_{3|1}(t)$ in the six-stage semi-Markov models under independent censoring.}
\label{Semi-Markov_I_F}
\end{figure}

\begin{figure}[H]
\centering
\includegraphics[scale=.70]{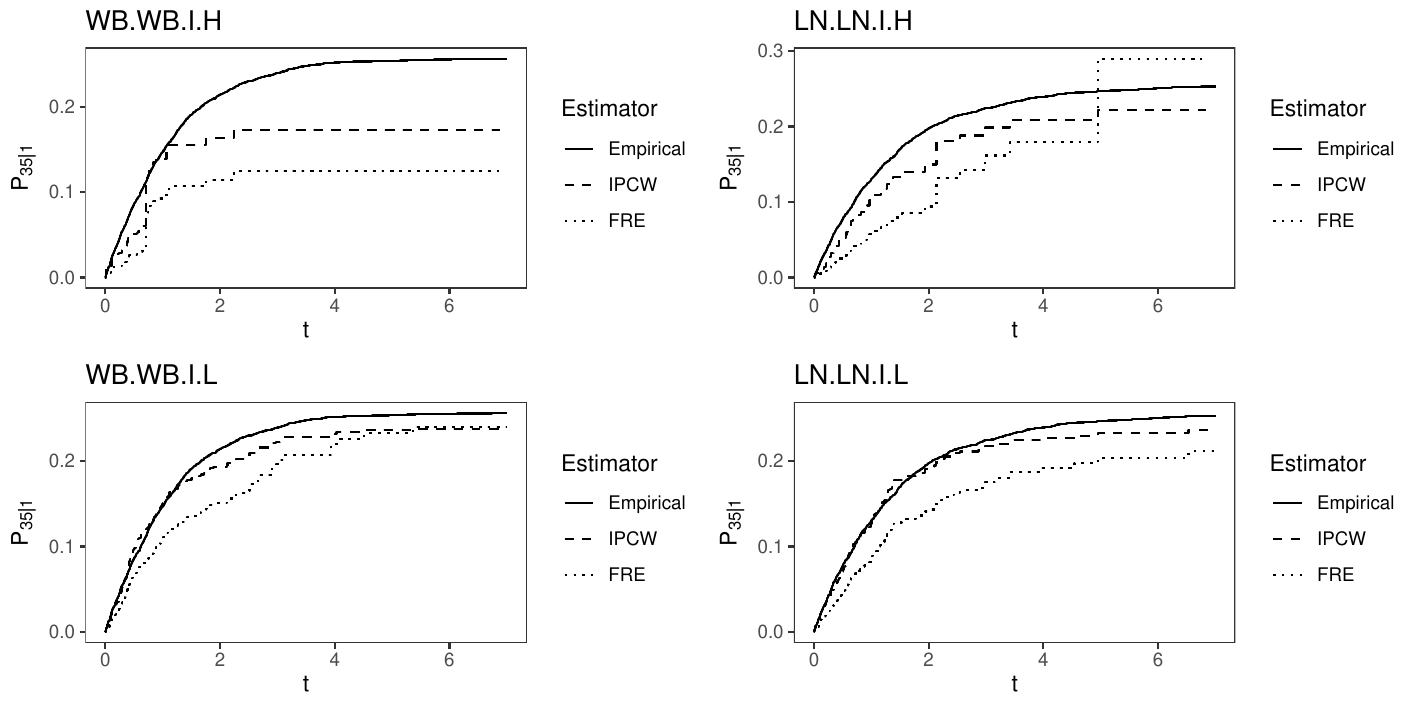}
\caption{Comparison of IPCW and FRE estimators (based on a sample size of 2000 under a specific simulation design) with the empirical true distribution (based on a sample size of 10000 under the simulation design with no censoring) for $P_{35|1}(t)$ in the six-stage Markov models under independent censoring.}
\label{Markov_I_P}
\end{figure}

\begin{figure}[H]
\centering
\includegraphics[scale=.70]{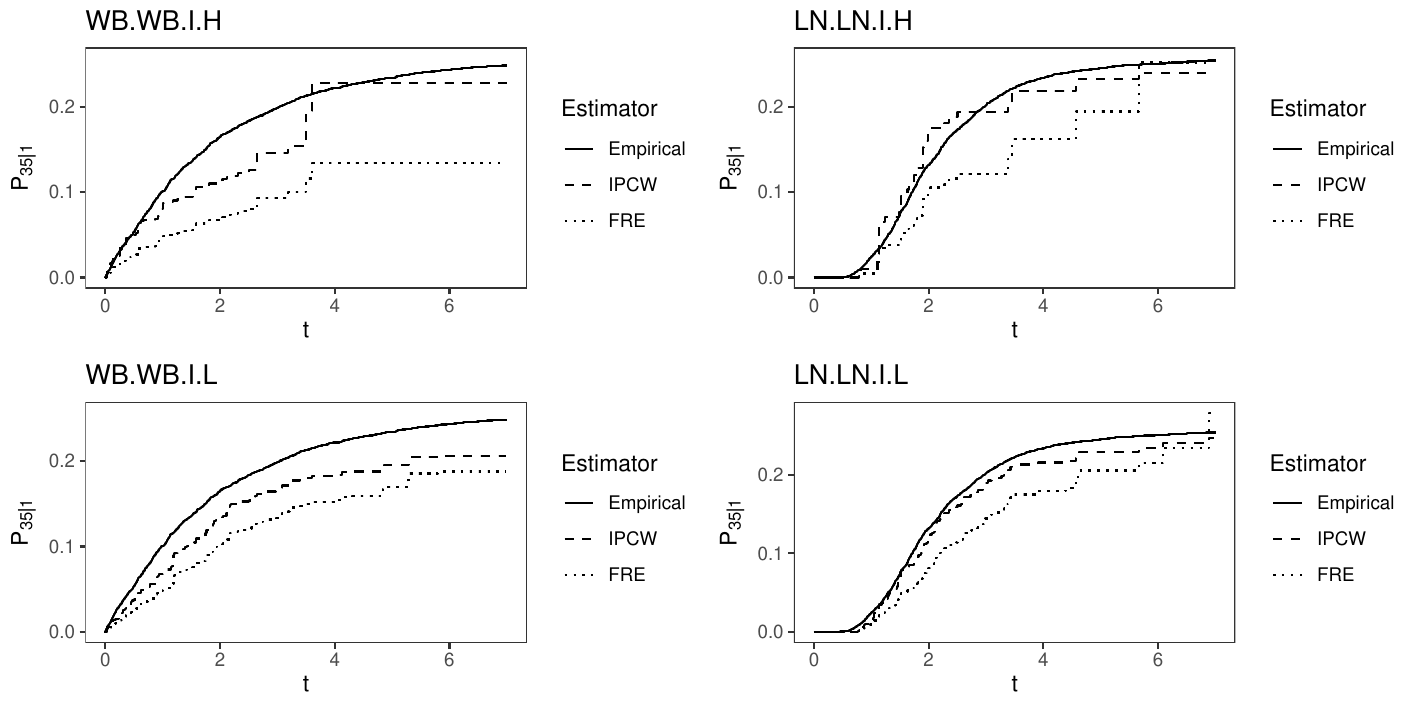}
\caption{Comparison of IPCW and FRE estimators (based on a sample size of 2000 under a specific simulation design) with the empirical true distribution (based on a sample size of 10000 under the simulation design with no censoring) for $P_{35|1}(t)$ in the six-stage semi-Markov models under independent censoring.}
\label{Semi-Markov_I_P}
\end{figure}

\begin{figure}[H]
\centering
\includegraphics[scale=.70]{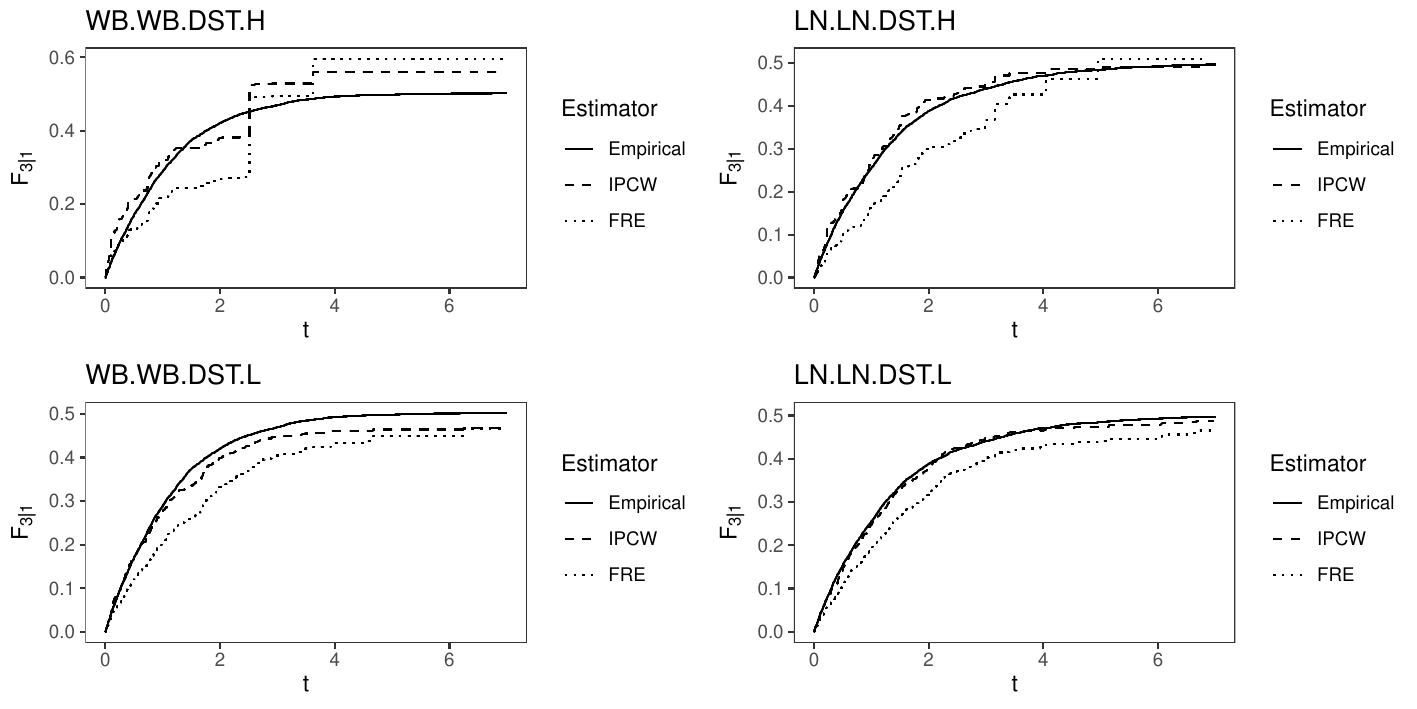}
\caption{Comparison of IPCW and FRE estimators (based on a sample size of 2000 under a specific simulation design) with the empirical true distribution (based on a sample size of 10000 under the simulation design with no censoring) for $F_{3|1}(t)$ in the six-stage Markov models under dependent censoring.}
\label{Markov_DST_F}
\end{figure}

\begin{figure}[H]
\centering
\includegraphics[scale=.70]{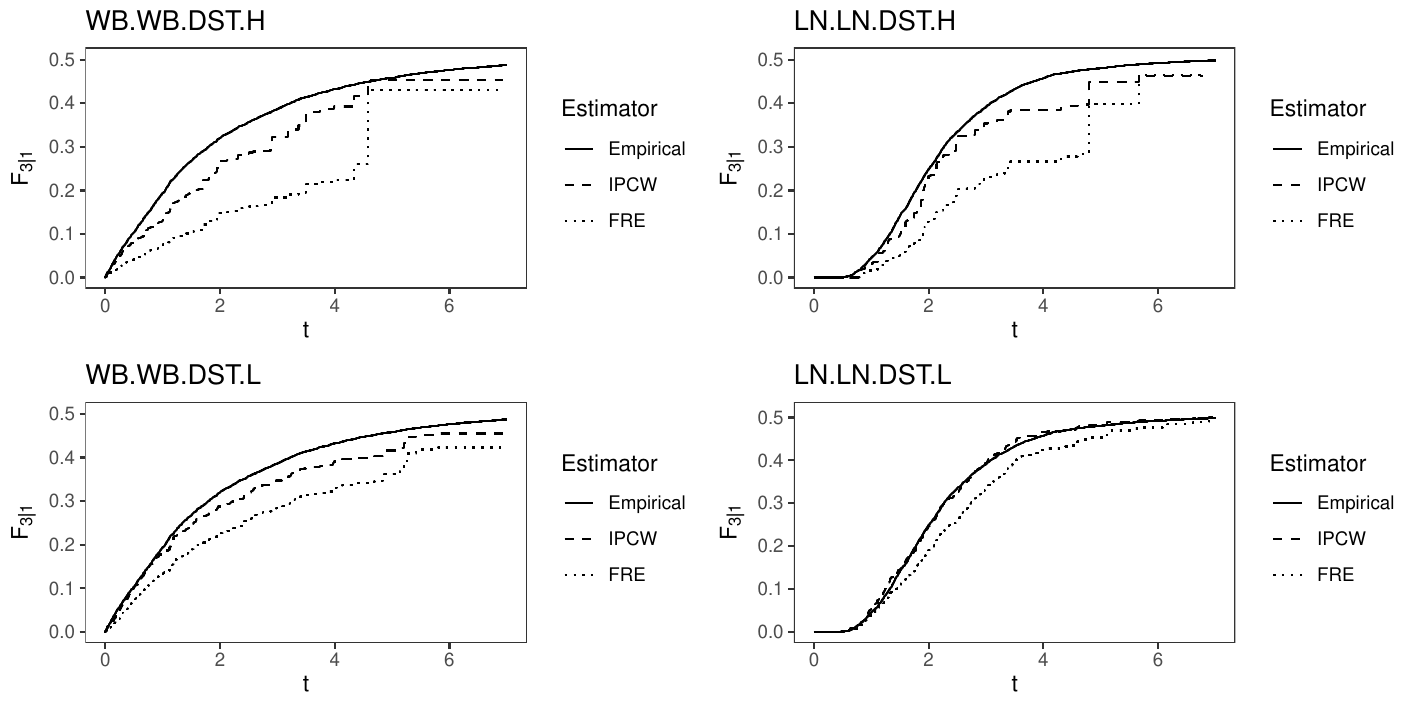}
\caption{Comparison of IPCW and FRE estimators (based on a sample size of 2000 under a specific simulation design) with the empirical true distribution (based on a sample size of 10000 under the simulation design with no censoring) for $F_{3|1}(t)$ in the six-stage semi-Markov models under dependent censoring.}
\label{semi-Markov_DST_F}
\end{figure}

\begin{figure}[H]
\centering
\includegraphics[scale=.70]{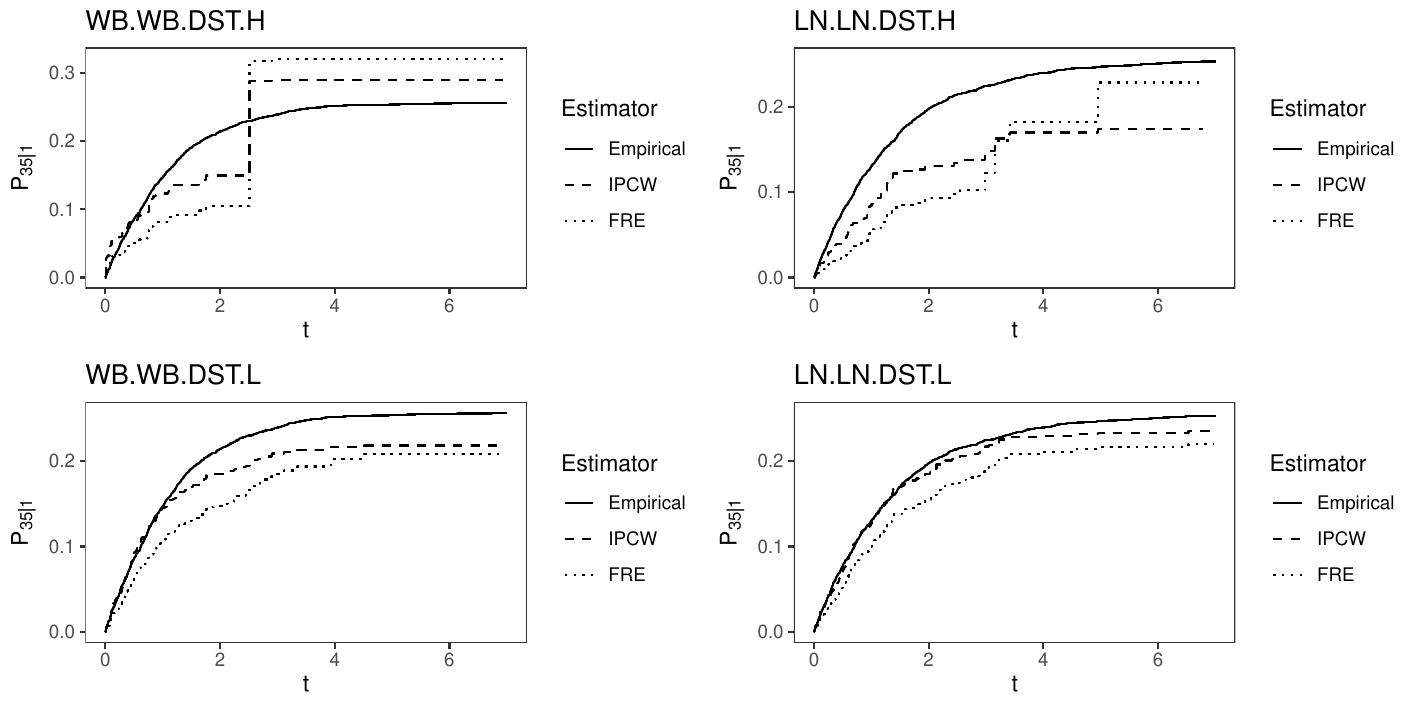}
\caption{Comparison of IPCW and FRE estimators (based on a sample size of 2000 under a specific simulation design) with the empirical true distribution (based on a sample size of 10000 under the simulation design with no censoring) for $P_{35|1}(t)$ in the six-stage Markov models under dependent censoring.}
\label{Markov_DST_P}
\end{figure}

\begin{figure}[H]
\centering
\includegraphics[scale=.70]{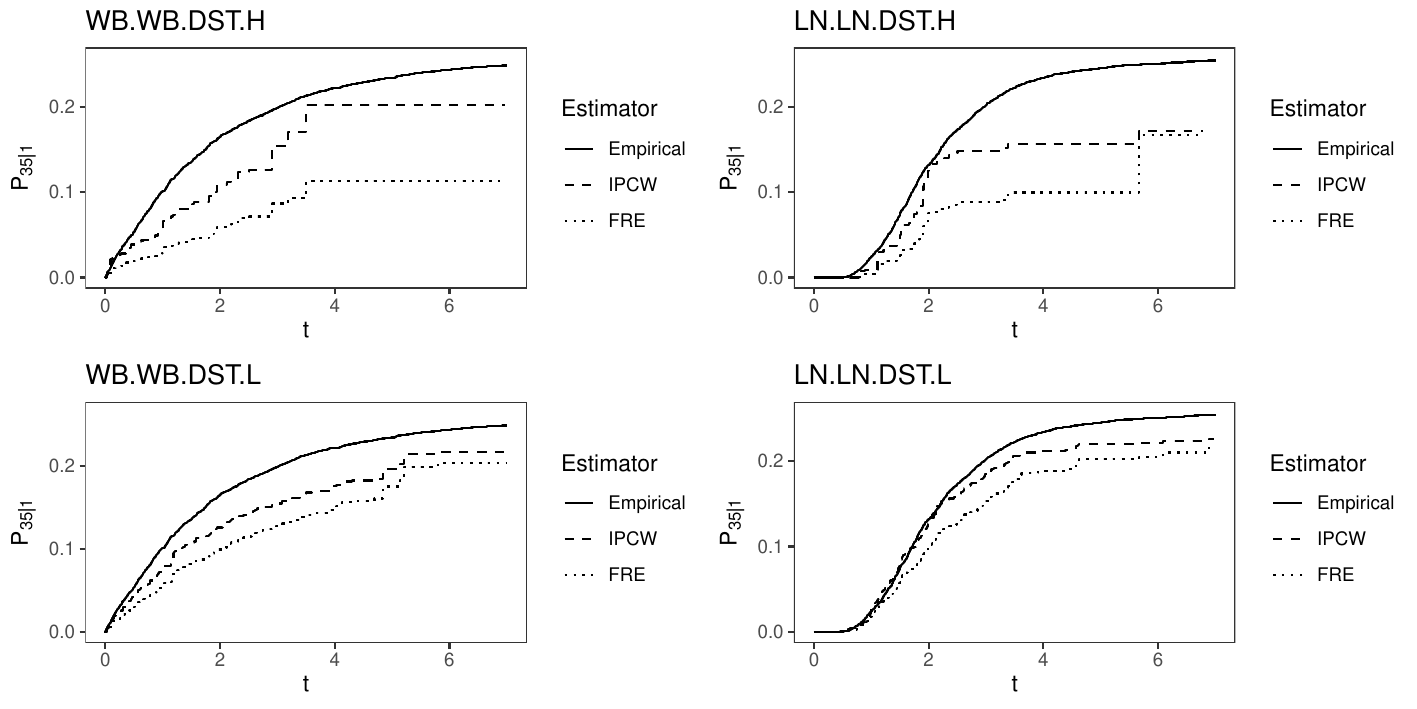}
\caption{Comparison of IPCW and FRE estimators (based on a sample size of 2000 under a specific simulation design) with the empirical true distribution (based on a sample size of 10000 under the simulation design with no censoring) for $P_{35|1}(t)$ in the six-stage semi-Markov models under dependent censoring.}
\label{Semi-Markov_DST_P}
\end{figure}

Web Figures \ref{Markov_I_F} -- \ref{Semi-Markov_DST_P} present the IPCW and FRE estimators for $F_{3|1}(t)$ (or $P_{35|1}(t)$), as well as their empirically true distributions obtained from a large simulated sample. In these figures, the solid curve represents the empirically true distribution of $F_{3|1}(t)$ (or $P_{35|1}(t)$), obtained from uncensored data simulations with a large sample size of 100,000 individuals. The dashed and dotted curves depict the IPCW and FRE estimates of $F_{3|1}(t)$ (or $P_{35|1}(t)$), respectively, with a sample size of 2000. The abbreviations in the top-left corner of each panel denote the simulation design. For example, ``WB.WB.DST.H'' (or ``LN.LN.I.L'') represents the scenario where waiting times follow a Weibull (or log-normal) distribution and censoring times also follow a Weibull (or log-normal) distribution under the dependent (or independent) censoring type with a high (or low) censoring rate scheme. 

\subsection*{Web Appendix D}
\subsubsection*{Breast cancer}\label{case: breast cancer}
\begin{figure}[H]
\centering
\includegraphics[scale=.50]{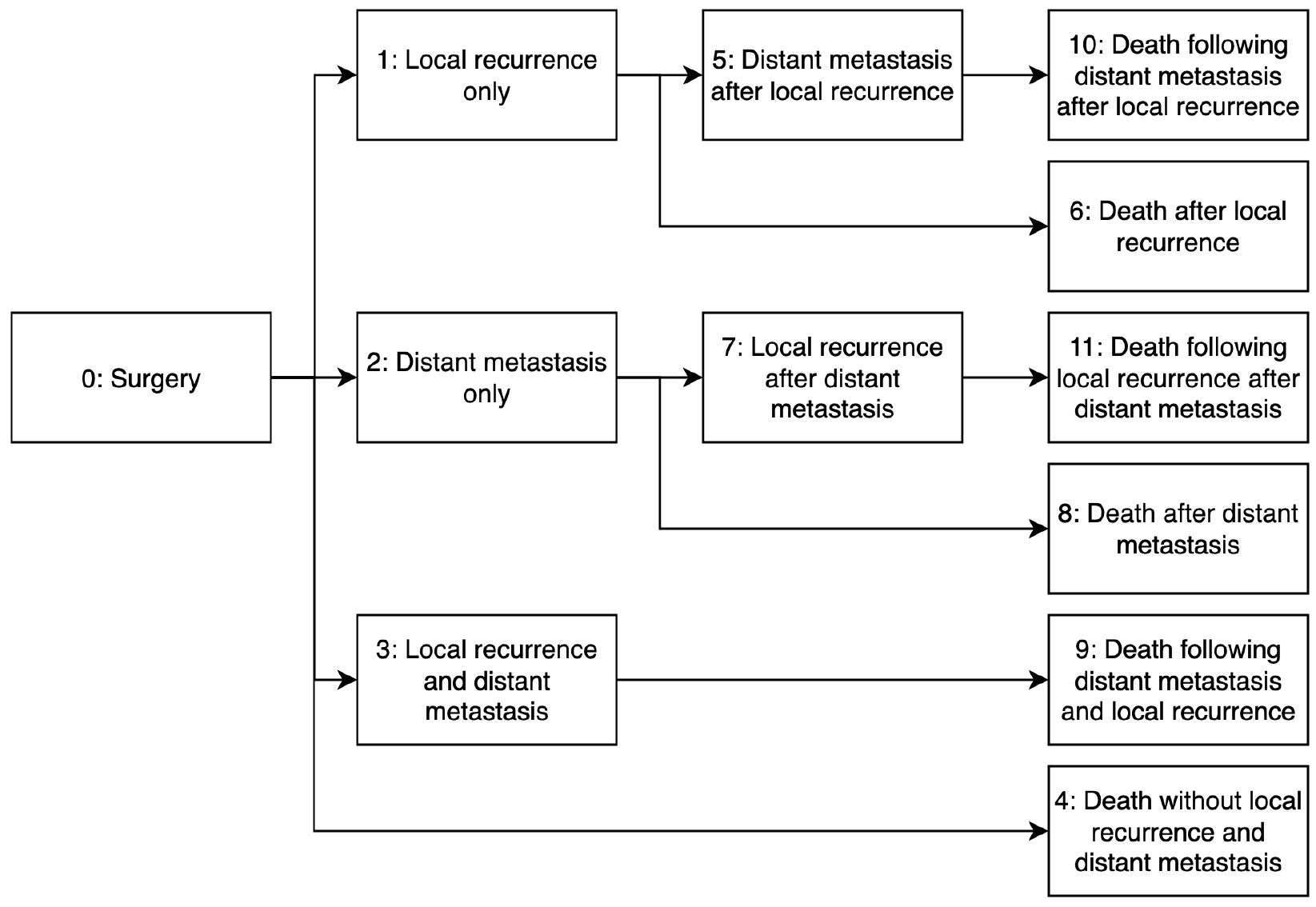}
\caption{Diagram of the twelve-stage survival model for the breast cancer data}
\label{eortc12stage}
\end{figure}

First, we apply the proposed estimation methods to breast cancer data from a trial conducted by the European Organization for Research and Treatment of Cancer (EORTC-trial 10854), involving 2793 patients with early breast cancer. A comprehensive description of the study is provided in Clahsen et al. (1996), and further findings are reported in Van der Hage et al. (2001). Stages are defined based on three post-surgery events: the onset of local recurrence, the onset of distant metastasis, and death. The resulting network of stages is shown in Web Figure \ref{eortc12stage}. This multi-stage model comprises twelve stages, including event-free and alive after surgery (stage 0); local recurrence following surgery (stage 1); distant metastasis following surgery (stage 2); simultaneous occurrence of local recurrence and distant metastasis following surgery (stage 3); death without local recurrence or distant metastasis (stage 4); distant metastasis after local recurrence (stage 5); death after local recurrence (stage 6); local recurrence after distant metastasis (stage 7); death after distant metastasis (stage 8); death after simultaneous local recurrence and distant metastasis (stage 9); death after distant metastasis following local recurrence (stage 10); and death after local recurrence following distant metastasis (stage 11). The time origin for the analysis was set as the time of surgery, indicating that all individuals started in stage 0. Transition times in the original dataset were right-censored, and follow-up times ranged from 12 to 6756 days.

Web Table \ref{eortc: transition} summarizes the observed number of transitions between stages in the breast cancer data. Diagonal elements in the table correspond to the number of censored observations in the stages, while off-diagonal elements count the observed transitions. There are no observed transitions starting from stages 4, 6, 8, 9, 10, and 11 since these stages are terminal stages, death, as no events can happen after death. Our goal is to estimate the stage waiting time distributions for different post-surgery events in patients who have just completed surgery. Specifically, we focus on $F_{1|0}$, $F_{2|0}$, $F_{3|0}$, $F_{5|0}$, and $F_{7|0}$, which are all conditional on the prior stage visit of stage 0. Regarding the cumulative incidence function conditional on the prior stage visit, we are interested in $P_{5\ 10|1}$ and $P_{7\ 11|2}$, where the event of interest is both terminal stage (death). 

\begin{table}[H]
\footnotesize
\centering
\setlength{\belowcaptionskip}{+0.2cm}
\caption{Observed transitions in the breast cancer data.}
\begin{tabular}{c ccc ccc ccc ccc}
\hline
& \multicolumn{12}{c}{To} \\
\cline{2-13}
From & 0    & 1    & 2    & 3    & 4    & 5    & 6    & 7    & 8    & 9    & 10    & 11    \\
\hline
0    & 1686 & 268  & 625  & 80   & 134  & 0    & 0    & 0    & 0    & 0    & 0     & 0     \\
1    & 0    & 143  & 0    & 0    & 0    & 89   & 36   & 0    & 0    & 0    & 0     & 0     \\
2    & 0    & 0    & 121  & 0    & 0    & 0    & 0    & 36   & 467  & 0    & 0     & 0     \\
3    & 0    & 0    & 0    & 12   & 0    & 0    & 0    & 0    & 0    & 0    & 0     & 0     \\
4    & 0    & 0    & 0    & 0    & 0    & 0    & 0    & 0    & 0    & 0    & 0     & 0     \\
5    & 0    & 0    & 0    & 0    & 0    & 16   & 0    & 0    & 0    & 0    & 72    & 0     \\
6    & 0    & 0    & 0    & 0    & 0    & 0    & 0    & 0    & 0    & 0    & 0     & 0     \\
7    & 0    & 0    & 0    & 0    & 0    & 0    & 0    & 3    & 0    & 0    & 0     & 33    \\
8    & 0    & 0    & 0    & 0    & 0    & 0    & 0    & 0    & 0    & 0    & 0     & 0     \\
9    & 0    & 0    & 0    & 0    & 0    & 0    & 0    & 0    & 0    & 0    & 0     & 0     \\
10   & 0    & 0    & 0    & 0    & 0    & 0    & 0    & 0    & 0    & 0    & 0     & 0     \\
11   & 0    & 0    & 0    & 0    & 0    & 0    & 0    & 0    & 0    & 0    & 0     & 0     \\
 \hline
\end{tabular}
\label{eortc: transition}
\end{table}

\begin{figure}[H]
\centering
\includegraphics[scale=.75]{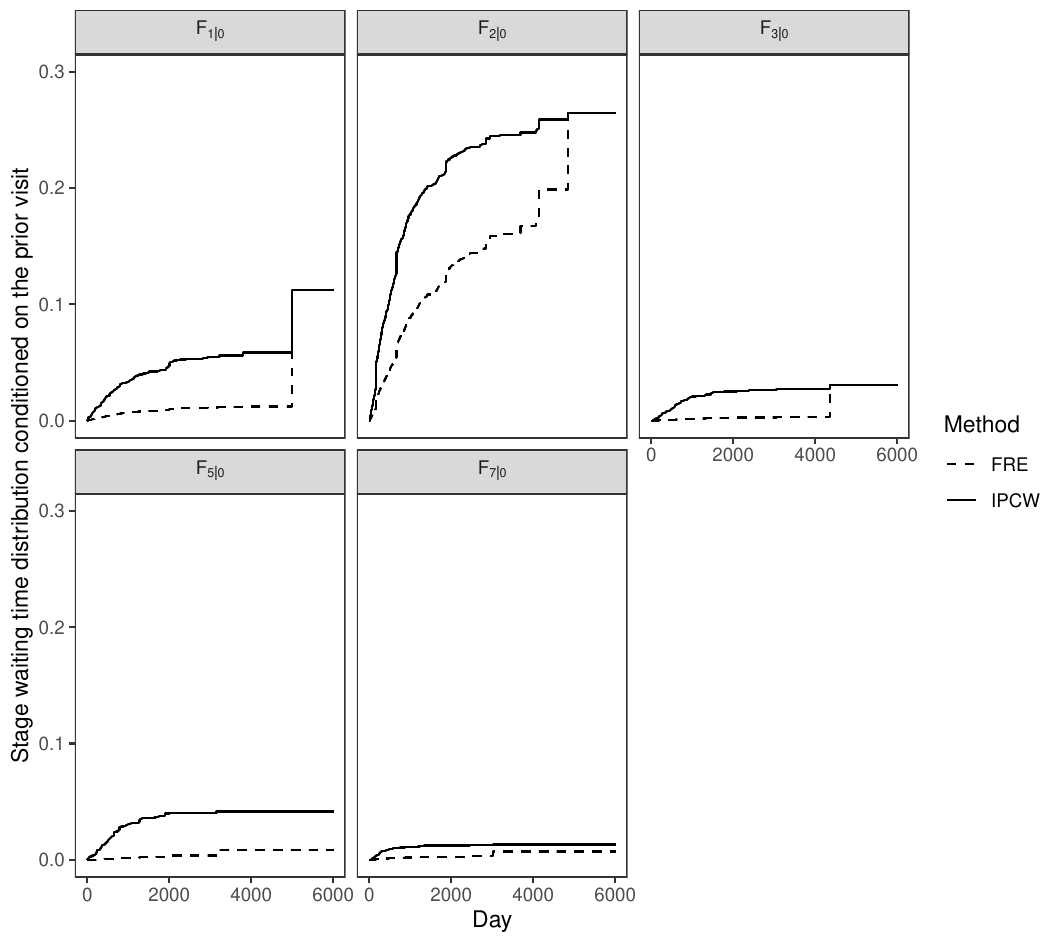}
\caption{IPCW and FRE estimates of the stage waiting time distributions conditional on the prior stage visit mentioned in Web Appendix D.}
\label{fig: eortc_f}
\end{figure}

\begin{figure}[H]
\centering
\includegraphics[scale=.58]{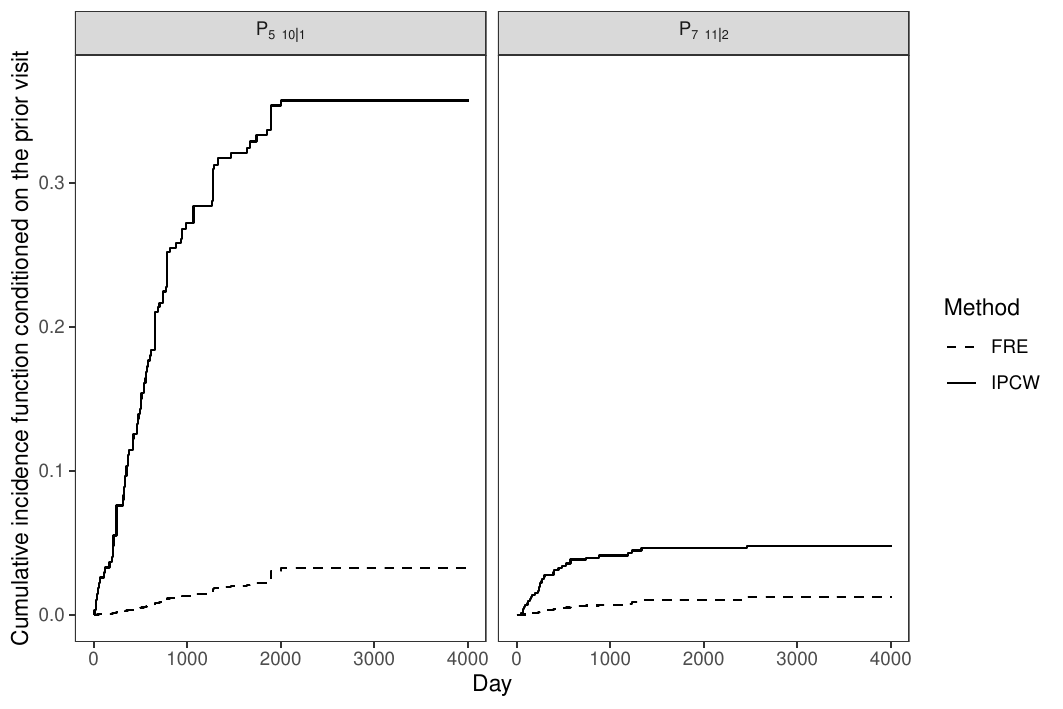}
\caption{IPCW and FRE estimates of the cumulative incidence functions conditional on the prior stage visit mentioned in Web Appendix D.}
\label{fig: eortc_p}
\end{figure}

Web Figures \ref{fig: eortc_f} and \ref{fig: eortc_p} show the IPCW and FRE estimations of the stage waiting time distributions and cumulative incidence functions conditional on the prior stage visit, mentioned above. A visual inspection of the solid lines in Web Figure \ref{fig: eortc_f} indicates some noticeable differences in these IPCW estimates of stage waiting time distributions conditional on the onset of surgery (stage 0). Specifically, patients after surgery (stage 0) are more likely to remain in distant metastasis only (stage 2) than in local recurrence only (stage 1) after the same amount of waiting time. This is followed by distant metastasis occurring after local recurrence (stage 5), the simultaneous occurrence of local recurrence and distant metastasis (stage 3), and local recurrence following distant metastasis (stage 7). Patients after surgery (stage 0) have the smallest probability of remaining in local recurrence after distant metastasis (stage 7) after the same amount of waiting time. Furthermore, from the solid lines in Web Figure \ref{fig: eortc_p}, it can be seen that a much higher percentage of patients who develop local recurrence only (stage 1) subsequently develop distant metastasis (stage 5) and eventually progress to death (stage 10) compared to those who develop distant metastasis only (stage 2), later develop local recurrence (stage 7), and ultimately progress to death (stage 11) within the same waiting time.

The FRE estimates, represented by the dashed lines in Web Figures \ref{fig: eortc_f} and \ref{fig: eortc_p}, exhibit trends similar to those of the IPCW estimates. However, the FRE estimates are generally more conservative, with lower estimated values. This difference can be attributed to the idea of a ``unique path'' used in constructing fractional observations and the substantial amount of censoring (see Web Table \ref{eortc: transition}), which may introduce greater estimation errors in the construction of at-risk sets. Specifically, Web Table \ref{eortc: transition} indicates that 1686 patients, approximately 60\% of the total number of patients, were censored at stage 0. The small proportion of observed transitions after stage 0 can impact the accuracy of the estimations. Consistent with the findings from the simulation studies, the FRE estimator tends to be less reliable than the IPCW estimator, particularly in the presence of a high censoring rate.

\newpage

\end{document}